\documentclass[a4paper,12pt]{article}
\usepackage[utf8]{inputenc}
\usepackage{multicol}
\usepackage{amsmath}
\usepackage{amssymb}

\usepackage{graphicx}
\usepackage[dvipsnames]{xcolor}
\usepackage{graphicx,times}             

\DeclareUnicodeCharacter{2212}{\textendash}

\usepackage{graphicx}
        \usepackage{dcolumn}
        \usepackage{bm}
        \usepackage[final]{hyperref}

        \usepackage{multirow}
        \usepackage[dvipsnames]{xcolor}
        \usepackage{comment}

\title{Supermassive Neutron Stars in Starobinsky Gravity with Causal Hybrid Stellar Matter}
\author{Muhammed Shafeeque,$^1$\footnote{m.shafeeque@iitg.ac.in} \ Arun Mathew$^2$\footnote{arun@cp.dias.ie} \ and Malay K. Nandy$^1$\footnote{mknandy@iitg.ac.in}\\
        $^1$\small\em Department of Physics, Indian Institute of Technology Guwahati, Guwahati 781039, India.\\
        $^2$\small\em Dublin Institute for Advanced Studies, Astronomy \& Astrophysics Section, Dublin D15 XR2R,
             Ireland.}

        \date{\normalsize (August 13, 2025)}

\begin{document}
\maketitle

\begin{abstract}
We investigate the stellar structure of neutron stars in the framework of Starobinsky gravity, characterized by a quadratic correction to the Einstein-Hilbert action, $f(R) = R + \alpha R^2$. In order to {\em preserve causality throughout\/} the star, we adopt a two-phase hybrid construction for the stellar matter, in which the core region consists of deconfined quark matter described by the MIT bag model, while the outer layers are composed of hadronic matter represented by unified equations of state such as SLy4, BSk20, and BSk21. Within this framework, we derive the modified field equations with static spherical symmetry, and numerically integrate the corresponding Tolman-Oppenheimer-Volkoff (TOV) equations with the chosen hybrid equations of state. Our analyses show that, unlike in general relativity, the Ricci scalar remains nonzero outside the stellar surface, and gradually falls to zero beyond 50 km, while the stellar surface remains within 10--12 km for the hybrid equations of state considered. This extended Ricci scalar profile arises from the extra degree of freedom (scalaron) inherent in Starobinsky gravity, which also contributes to the gravitational mass outside the star, causing the ADM mass measured at infinity to exceed the stellar mass at the surface. Nevertheless, the key physical relationships, such as the mass-central density and mass-radius curves, remain consistent with what is expected physically. Notably, we find that the maximum stable mass of neutron stars increases with the Starobinsky parameter $\alpha$, with the combined MIT-BSk21 model supporting an ADM mass of up to $2.07 \, M_\odot$ for $\alpha = 10\,r_g^2$. This theoreical limit for a nonrotating neutron star suggests that a rotating configuration could reach mass thresholds in the range of $2.48$ to $2.59 \, M_\odot$, considering that rapid rotation can enhance the maximum mass of a neutron star by approximately 20--25\% compared to its non-rotating counterpart. These theoretical predictions align well with the observed masses of the heavy millisecond pulsars exceeding 2 M$_\odot$, highlighting the potential of $f(R)$ gravity to support massive neutron stars beyond the general relativistic limit while preserving causality.\\
{\em Keywords}: Modified gravity; $f(R)$ gravity; Quark matter; Unified hadronic matter; Causal stellar models
\end{abstract}


\section{Introduction}\label{sec_intro}

One of the notable predictions of general relativity (GR) is the existence of neutron stars \cite{landau1932,tolman,opp_volk}, where gravitational effects are extreme. The discovery of pulsars \cite{Giacconi1962,HEWISH1968}, later identified by Gold \cite{GOLD1968,GOLD1969} as rapidly rotating neutron stars, confirmed this prediction. Additional support came in 1975 with the discovery of the first binary neutron star system by Hulse and Taylor \cite{Hulse1975}, which also provided indirect observational evidence for gravitational waves. For decades following these discoveries, all measured pulsar masses remained well below $2$ M$_\odot$, consistent with GR-based expectations.

A major shift occurred after 2010 with the detection of pulsars whose masses approached or exceeded $2$ M$_\odot$. In 2010, Demorest et al. \cite{Demorest2010} reported MSP J1614-2230, with an updated mass measurement by Fonseca et al. \cite{Fonseca_2016} of $1.928\pm0.017$ M$_\odot$. In 2013, Antoniadis et al. \cite{Antoniadis2013} measured MSP J0348+0432 to be $2.01\pm0.04$ M$_\odot$. Subsequent observations have revealed even more massive pulsars, including PSR J2215+5135 with $2.27^{+0.15}_{-0.17}$ M$_\odot$ \cite{Linares2018}, MSP J0740+6620 with $2.14^{+0.10}_{-0.09}$ M$_\odot$ \cite{Cromartie2020}, and the detection of PSR J0952-0607 by Bassa et al. \cite{PSRJ095206071}, whose mass was determined to be $2.35\pm0.17$ M$_\odot$ by Romani et al. \cite{PSRJ095206072}.

Gravitational wave astronomy has provided further evidence for high-mass neutron stars. The binary neutron star merger GW170817 \cite{GW170817}, detected by LIGO, involved component masses between $1.0$ and $1.89$ M$_\odot$, with a total system mass of $2.73^{+0.04}_{-0.01}$ M$_\odot$ \cite{GW170817a}. Furthermore, GW190425 \cite{GW190425} indicated component masses in the range $1.15$-$2.52$ M$_\odot$, with a combined mass of $3.4^{+0.3}_{-0.1} $ M$_\odot$. Another remarkable event, GW190814 \cite{GW190814}, was interpreted as a merger between a $2.50$-$2.67 $ M$_\odot$ compact object and a black hole of $22.2$-$24.3 $ M$_\odot$. The nature of the lighter object remains uncertain, lying in the so-called mass gap---too massive to be a standard neutron star yet too light to be a typical black hole---making it a strong candidate for an ultra-massive neutron star.

The existence of such massive neutron stars poses significant theoretical challenges, particularly concerning their internal composition and the properties of matter at supernuclear densities. These observations place stringent constraints on the equation of state, which must be sufficiently stiff to support stars of $\sim 2$ M$_\odot$ \cite{massradii}. One strategy to achieve the necessary stiffness is to suppress the presence of hyperons in dense nuclear matter \cite{nohyp1,nohyp2,fermiliqeos}. However, theoretical models suggest that hyperons are likely to form under extreme densities, softening the equation of state and reducing the maximum stable mass \cite{hyperon1,hyperon2}.

Numerous equation of state models for neutron-rich matter have been proposed, incorporating multi-body interactions and different phenomenological approaches \cite{SK0,SK1,FPS1,FPS0,SLy_Chabanat_Haensel_1,SLy_Chabanat_Haensel2,RMF1,RMF2,RMF3,EFT,APR,fermiliqeos}. While some of these are stiff enough to yield $M_{\rm max}\sim 2$ M$_\odot$ or higher, certain models violate causality at high densities. Achieving maximal masses around $2$ M$_\odot$ often requires central densities on the order of $\rho_c \sim 10\rho_{\rm nuc}$, where $\rho_{\rm nuc} = 2.67 \times 10^{14}$ g cm$^{-3}$ is the nuclear saturation density. At such extreme densities, quark deconfinement is expected to occur in the stellar core \cite{hydrostatic_MIT_Naok,quark_stars_Ivanenko,SHURYAK}.

Extended theories of gravity have also been explored as a means to achieve neutron star masses above $2$ M$_\odot$ \cite{Antonio,Gonzalo}. In this paper, we focus on the Starobinsky $f(R)$ gravity model \cite{STAROBINSKY1980}, given by $f(R) = R + \alpha R^2$. Treating the $\alpha R^2$ term as a perturbation, Arapoğlu et al. \cite{Arapo_lu2011} found that, for certain equations of state, the maximum mass could approach $\sim 2$ M$_\odot$ only for negative $\alpha$. By contrast, Yazadjiev et al. \cite{Yazadjiev2014} solved for stellar equilibrium configurations non-perturbatively using an equivalent scalar-tensor formulation, avoiding the unphysical behaviour seen in perturbative treatments, such as a decrease in enclosed mass within the stellar interior \cite{Orellana2013}.

Astashenok et al. \cite{Astashenok2017} extended the non-perturbative analysis of $R+\alpha R^2$ gravity to both neutron and quark stars. For positive $\alpha$, they found that the Ricci scalar remains finite at the surface and decays exponentially outside the star, adding an external gravitational mass component to the total observed mass. This feature influences observable effects such as gravitational redshift.

The viability of $f(R)$ gravity in high-curvature regimes has been debated, with some works \cite{Kobayashi,AndreiFrolov,Kobayashi2} reporting the formation of curvature singularities inside highly compact stars. However, Babichev and Langlois \cite{BabichevLanglois} demonstrated that stable, relativistic stars can be obtained numerically in $f(R)$ gravity using polytropic equations of state, provided the condition $\rho c^2 - 3P > 0$ is satisfied throughout the stellar interior. This result was later confirmed in \cite{Mass_rad} using unified equations of state. Nonetheless, the same condition restricts the stiffness of the equation of state, thereby limiting the maximum stable mass achievable in $f(R)$ gravity.

From the above discussion, it follows that the condition $\rho c^2 - 3P > 0$ constrains the equation of state in such a way that it remains causal, avoiding superluminal sound speeds. Under this restriction, an appropriately chosen $f(R)$ gravity model may still accommodate neutron stars with maximum stable masses exceeding $2$~M$_\odot$. This is plausible since the $f(R)$ model, such as $R + \alpha R^2$ gravity, possesses a tunable parameter $\alpha$. Motivated by this, in the present paper we explore hybrid equations of state \cite{Shafeeque2023} in which the stellar core is composed of deconfined quark matter, while the outer layers are formed of hadronic matter. The presence of the quark core is expected to maintain the equation of state within the causal limit at high densities.

In the present analysis, the quark phase is modeled using the MIT bag model \cite{MIT_jeff,MIT_Yu.A.Simonov}, while the hadronic phase is described by unified equations of state such as SLy4, BSk20 and BSk21 \cite{SLy_Haensel1_Potekhin,bsk_Potekhin}. This two-phase construction ensures that the transition from hadronic matter to quark matter preserves causality {\em throughout\/} the stellar interior. As summarized earlier in Table~1 in \cite{Shafeeque2023}, these hybrid equations of state yield maximum masses marginally below $2$~M$_\odot$ based on GR, with the MIT-BSk21 combination coming closest to this threshold.

The remainder of this paper is structured as follows. Section~\ref{ch3_action_field_eqn} introduces the action for $f(R)$ gravity, supplemented by a general matter action, and derives the corresponding modified field equations for the Starobinsky $f(R)$ model. In Section~\ref{ch3_sec_TOV}, we specialize to a static, spherically symmetric spacetime and extract the differential equations governing the metric potentials and the Ricci scalar from the modified field equations. Section~\ref{ch3_sec_eq_state} presents the complete set of equations of state employed in this analysis, while Section~\ref{ch3_sec_ic} outlines the initial conditions used for the numerical integration. The results of these integrations, including the radial profiles of pressure, Ricci scalar, and mass function, as well as the dependence of the ADM mass on central density and stellar radius, are discussed in Section~\ref{ch3_result}. Finally, Section~\ref{ch3_conc} provides concluding remarks.

\section{Action and field equations}\label{ch3_action_field_eqn} 

The most general form of the action for $f(R)$ gravity is given by
\begin{equation}\label{ch3_fR_action}
 \mathcal{S}(\Psi,g)=\int d^{4}x\,\sqrt{-g}\left\{ \frac{c^{3}}{16\pi G}f\left(R\right)+\mathcal{L}_{m}(\Psi,g)\right\},
\end{equation}
where $R$ is the Ricci scalar, $g$ is the determinant of the metric tensor $g_{\mu\nu}$, $G$ is the gravitational constant, $c$ is the velocity of light, and $\mathcal{L}_m(\Psi,g)$ is the matter Lagrangian with $\Psi$ being the matter field. The energy-momentum tensor is obtained from the matter Lagrangian using
\begin{equation}\label{ch3_eq_em_tensor}
T_{\mu\nu}=cg_{\mu\nu}\mathcal{L}_{m}-2c\frac{\delta\mathcal{L}_{m}}{\delta g^{\mu\nu}}.
\end{equation}

In the metric formalism, the action (\ref{ch3_fR_action}) is varied with respect to the metric tensor $g_{\mu\nu}$ to obtain the 
modified field equations in the $f(R)$ gravity framework.  Demanding the extremality condition $\delta \mathcal{S}=0$ gives the modified field equations as
\begin{equation}\label{ch3_FE_fR}
 f_{R}R_{\mu\nu}-\frac{1}{2}g_{\mu\nu}f+\left(g_{\mu\nu}\Box-\nabla_{\mu}\nabla_{\nu}\right)f_{R}=\kappa\,T_{\mu\nu},
\end{equation}
where $\kappa=\frac{8\pi G}{c^{4}}$, and $f_{R}=\frac{\partial f}{\partial R}$.

It can be shown that the energy-momentum conservation $\nabla_\mu T^{\mu}_{\ \nu}=0$ is satisfied in $f(R$) gravity. Using Bianchi identity and general properties of tensors, we obtain the covariant derivative of the left-hand side of the modified field equations (\ref{ch3_FE_fR}),
\begin{equation}
 \nabla_{\mu}\left[f_{R}R^{\mu}_{\ \nu}-\frac{1}{2}\delta_{\ \nu}^{\mu}f+\left(\delta_{\ \nu}^{\mu}\Box-\nabla^{\mu}\nabla_{\nu}\right)f_{R}\right]=0,
\end{equation}
satisfiying the conservation of energy-momentum tensor.

In the present work, we use the Starobinky $f(R)$ model \cite{STAROBINSKY1980}, with
\begin{equation}\label{ch3_starobisnky_fR}
 f(R)=R+\alpha R^2,
\end{equation}
where $\alpha$ is a coupling parameter, having the dimension of $[L^{2}]$. Using this form of $f(R)$ in (\ref{ch3_FE_fR}), we obtain the modified field equations in the Starobinky $f(R)$ model as
\begin{equation}\label{ch3_FE_starobinsky}
 G_{\mu\nu}+2\alpha R\left(R_{\mu\nu}-\frac{1}{4}g_{\mu\nu}R\right)+2\alpha\left(g_{\mu\nu}\Box-\nabla_{\mu}\nabla_{\nu}\right)R=\kappa\,T_{\mu\nu},
\end{equation}
where $G_{\mu\nu}$ is the Einstein tensor, defined as $G_{\mu\nu}=R_{\mu\nu}-\frac{1}{2}g_{\mu\nu}R$. Taking the trace of equation (\ref{ch3_FE_starobinsky}) gives
\begin{equation}\label{ch3_staro_trace_eqn}
 6\alpha\Box R=R+\kappa\,T,
\end{equation}
which is a clear departure from the trace of Einstein field equations, $R=-\kappa\,T$. This result (\ref{ch3_staro_trace_eqn}) has significant impact in the stellar structure of the neutron star, which is discussed in detail in the following sections.

\section{Spherically symmetric solution}\label{ch3_sec_TOV}

We assume the star to be static, spherically symmetric and non-rotating. Therefore, the most general metric that describes the spacetime is given by
\begin{equation}\label{ch3_metric}
 ds^{2}=-e^{\nu(r)}\,c^{2}\,dt^{2}+e^{\lambda(r)}\,dr^{2}+r^{2}d\theta^{2}+r^{2}\sin^{2}\theta \,d\varphi^{2},
\end{equation}
where $\lambda(r)$ and $\nu(r)$ are the metric potentials. 

The matter present inside a static neutron star is a perfect fluid. The energy-momentum tensor of a perfect fluid is given by \cite{LANDAU197566}
\begin{equation}\label{ch3_PF}
 T_{\mu\nu}=\frac{\left(\varepsilon+P\right)}{c^{2}}u_{\mu}u_{\nu}+g_{\mu\nu}P,
\end{equation}
where $\varepsilon=\rho c^2$ is the energy density, $P$ is the pressure, and $u^\mu$ is the four-velocity, with $u^\mu u_\nu=-c^2$.

Using the metric (\ref{ch3_metric}), the $tt$ and $rr$ components of the modified field equations (\ref{ch3_FE_starobinsky}) yield the equations for the metric potentials:
\begin{equation}\label{ch3_lambda_DE}
 \lambda^{\prime}=\frac{1-e^{\lambda}}{r}+\frac{re^{\lambda}}{6}\left(\frac{2R+3\alpha R^{2}}{1+2\alpha R}\right)-\left(\frac{\alpha rR^{\prime}}{1+2\alpha R}\right)\nu^{\prime}+\frac{\kappa re^{\lambda}}{3}\left(\frac{T+3\varepsilon}{1+2\alpha R}\right)
\end{equation}
and
\begin{equation}\label{ch3_nu_DE}
 \gamma\nu^{\prime}=\frac{e^{\lambda}-1}{r}+\frac{\kappa re^{\lambda}P}{1+2\alpha R}-\frac{r\alpha e^{\lambda}R^{2}}{2\left(1+2\alpha R\right)}-\frac{4}{r}\left(\frac{\alpha rR^{\prime}}{1+2\alpha R}\right),
\end{equation}
where $\gamma=1+\frac{\alpha rR^{\prime}}{1+2\alpha R}$.

On the other hand, using the metric (\ref{ch3_metric}) and the trace of the energy-momentum tensor (\ref{ch3_PF}), the trace equation (\ref{ch3_staro_trace_eqn}) leads to
\begin{equation}\label{ch3_Ricci_DE}
R^{\prime\prime}=\left(\frac{\lambda^{\prime}-\nu^{\prime}}{2}-\frac{2}{r}\right)R^{\prime}+\frac{\kappa}{6\alpha}e^{\lambda}T+\frac{e^{\lambda}R}{6\alpha},
\end{equation}
where $T=3P-\varepsilon$ is the trace of $T_{\mu\nu}$. 

Moreover, using the energy-momentum tensor (\ref{ch3_PF}) in the conservation of energy-momentum tensor, $\nabla_\mu T^{\mu}_{\ \nu}=0$, we get
\begin{equation}\label{ch3_TOV_dP}
P^{\prime}=-\frac{\nu^{\prime}}{2}\left(P+\varepsilon\right),
\end{equation}
which is the well-known Tolman-Oppenheimer-Volkoff equation \cite{tolman,opp_volk}. 

The cumulative mass $m(r)$, mass enclosed within a radial distance $r$, is obtained from the metric potential $\lambda(r)$ as
\begin{equation}\label{ch3_lambdamass}
 m(r)=\frac{c^{2}}{2G}r\left(1-e^{-\lambda(r)}\right).
\end{equation}

In the exterior of the star, the energy-momentum tensor $T_{\mu\nu}$ vanishes. We expect the pressure $P=0$ at the stellar surface as there is no matter beyond the surface. Thus, the trace of the energy-momentum tensor vanishes ($T=0$) in the exterior of the star. Consequently, from equation (\ref{ch3_Ricci_DE}), we have
\begin{equation}\label{ch3_Ricci_DE_exterior-a}
 R^{\prime\prime}=\left(\frac{\lambda^{\prime}-\nu^{\prime}}{2}-\frac{2}{r}\right)R^{\prime}+\frac{e^{\lambda}R}{6\alpha}.
\end{equation}
Noting that the last term in the above equation dominates as the coupling parameter $\alpha\rightarrow0$, we obtain $R=0$ as its solution, which is equivalent to general relativity. It is therefore evident that, in the weak gravity limit, we recover the general relativistic solution.

It is therefore safe to demand $\mu(r),\nu(r)\rightarrow0$ and $\mu^\prime(r),\nu^\prime(r)\rightarrow0$  as $r\rightarrow\infty$, in order to obtain an asymptotically flat spacetime. Consequently, equation (\ref{ch3_Ricci_DE_exterior-a}) in the asymptotic limit $r\rightarrow\infty$ takes the form
\begin{equation}\label{ch3_Ricci_DE_exterior}
 R^{\prime\prime}+\frac{2}{r}R^{\prime}-\frac{R}{6\alpha}=0,
\end{equation}
which has the general solution
\begin{equation}\label{ch3_Ricci_asym}
 R(r)=\frac{c_{1}}{r}e^{-\frac{r}{\sqrt{6\alpha}}}+\frac{c_{2}}{2r}\sqrt{6\alpha}e^{\frac{r}{\sqrt{6\alpha}}}.
\end{equation}
The constant of integration $c_2$ must vanish since $R\rightarrow0$ in the asymptotic limit $r\rightarrow\infty$, giving
\begin{equation}\label{ch3_Ricci_asym1}
 R(r)=\frac{c_{1}}{r}e^{-\frac{r}{\sqrt{6\alpha}}}.
\end{equation}
which approaches zero exponentially for values of $\alpha>0$, faster than $\frac{1}{r}$, as $r\rightarrow\infty$. However, when $\alpha<0$, the solution is oscillatory in nature, which is unphysical.

\section{Equations of State}\label{ch3_sec_eq_state}

The Tolman-Oppenheimer-Volkoff (TOV) equations must be supplemented by an equation of state in order to provide a closed system of equations that can be solved for the internal structure of a neutron star. This equation of state is determined by the properties of nuclear matter under extreme conditions. In the core of a neutron star, the matter experiences enormous pressure, with central densities reaching $\rho_c \sim 10^{15}$ g cm$^{-3}$ for configurations near the maximum mass. At such densities, the nuclear matter becomes highly degenerate, and its thermodynamic behavior becomes largely independent of temperature, except in a thin outer layer \cite{SLy_Haensel1_Potekhin}. 

In this work, we employ the combined set of equations of state provided in the Ref.~\cite{Shafeeque2023}. The core region is expected to contain deconfined quark matter, which we model using the MIT bag model \cite{MIT_jeff, MIT_Yu.A.Simonov}. The outer layers, composed of hadronic matter, are described using unified equations of state such as SLy \cite{SLy_Haensel1_Potekhin}, BSk20, and BSk21 \cite{bsk_Potekhin}.

 \begin{figure}
 \centering
 \includegraphics[width=.8\textwidth, height=8cm]{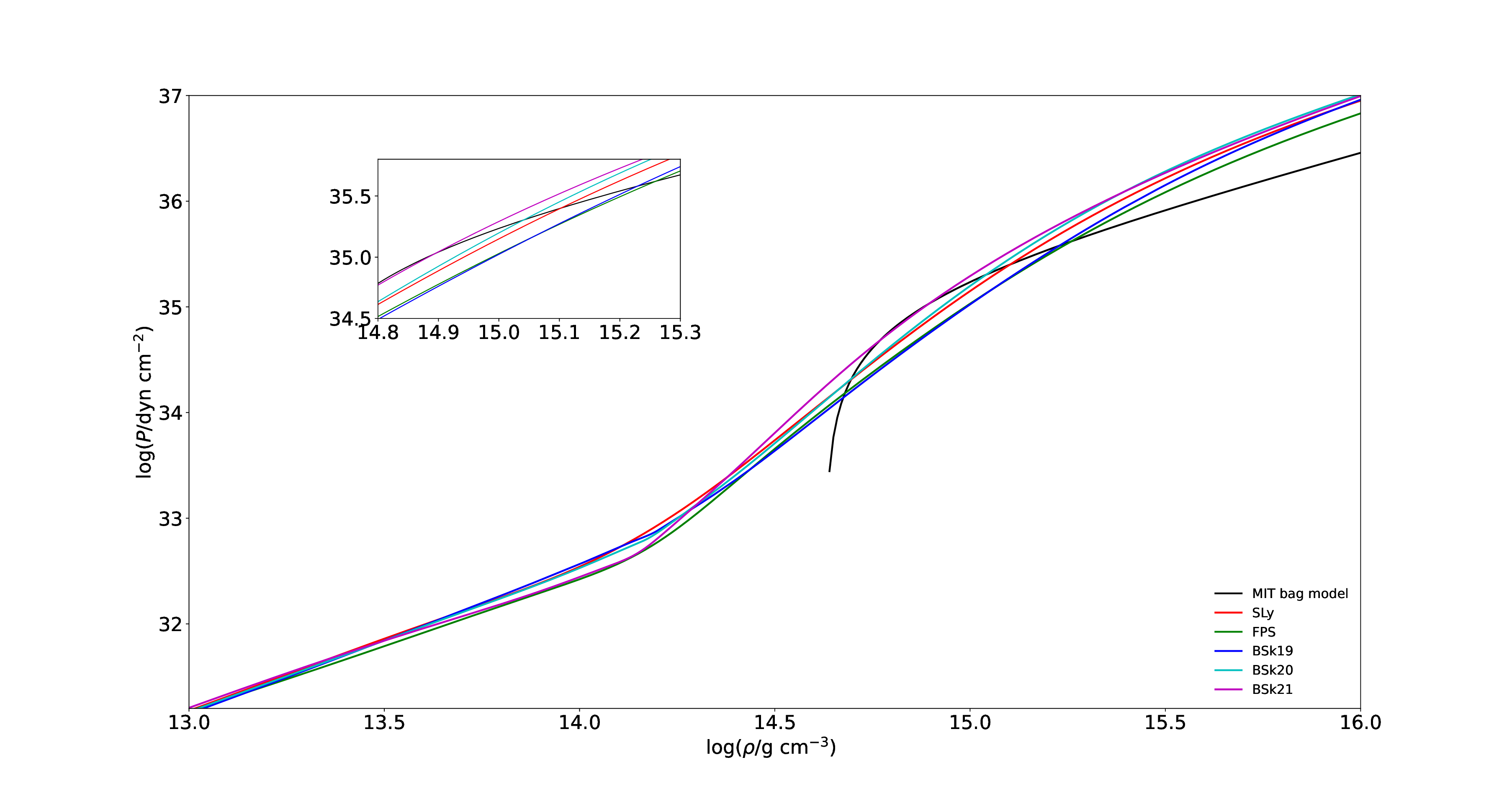}
 \caption{ \footnotesize Plots for different unified equations of state expressed by equations  \ref{SLy-FPS_eqn} and  \ref{bsk_eqns}, and the MIT bag model given by equation  \ref{mit_bag_model}. The inset shows the crossover points between the MIT bag model and the unified equations of state.}
 \label{fig.eos}
\end{figure}

To ensure smooth and continuous representations of the unified equations of state, we use parameterized expressions for the pressure $P$ as a function of density $\rho$. In the case of the SLy unified equation of state, the parameterization is given by \cite{SLy_Haensel1_Potekhin}:
\begin{equation}\label{SLy-FPS_eqn}
 \zeta=\frac{a_{1}+a_{2}\xi+a_{3}\xi^{3}}{(a_{4}\xi+1)\left(e^{a_{5}(\xi-a_{6})}+1\right)}+\frac{a_{7}+a_{8}\xi}{e^{a_{9}(a_{10}-\xi)}+1}+\frac{a_{11}+a_{12}\xi}{e^{a_{13}(a_{14}-\xi)}+1}+\frac{a_{15}+a_{16}\xi}{e^{a_{17}(a_{18}-\xi)}+1},\end{equation}
where $\zeta=\log\,(P/P_{\rm ref})$ with $P_{\rm ref}=1.0$ dyne cm$^{-2}$, and $\xi=\log\,(\rho/\rho_{\rm ref})$ with $\rho_{\rm ref}=1.0$ g cm$^{-3}$. The eighteen parameters $a_i$ are given in Ref.~\cite{SLy_Haensel1_Potekhin}.

Furthermore, the pressure $P$ as a function of density $\rho$ for the BSk20 and BSk21 unified equations of state is also provided in a parameterized form, 
\begin{equation}\label{bsk_eqns}
\begin{split}\zeta= & \frac{a_{1}+a_{2}\xi+a_{3}\xi^{3}}{\left(a_{4}\xi+1\right)\left(e^{a_{5}(\xi-a_{6})}+1\right)}+\frac{a_{7}+a_{8}\xi}{e^{a_{9}(a_{6}-\xi)}+1}+\frac{a_{10}+a_{1}\xi}{e^{a_{12}(a_{13}-\xi)}+1}\\
 & \ \ +\frac{a_{14}+a_{15}\xi}{e^{a_{16}(a_{17}-\xi)}+1}+\frac{a_{18}}{\left[a_{19}(\xi-a_{20})\right]^{2}+1}+\frac{a_{21}}{\left[a_{22}(\xi-a_{23})\right]^{2}+1},
\end{split}
\end{equation}
as given in \cite{bsk_Potekhin}, with $\zeta=\log\,(P/P_{\rm ref})$ and $\xi=\log\,(\rho/\rho_{\rm ref})$. The twenty-three parameters $a_i$ are also specified in Ref.  \cite{bsk_Potekhin}.

For the deconfined quark core, we adopt the equation of state described by the MIT bag model \cite{MIT_jeff, MIT_Yu.A.Simonov},
\begin{equation}\label{mit_bag_model}
P = k(\rho c^2 - 4B),
\end{equation}
where $B$ is the bag constant, and the parameter $k$ depends on the mass of the strange quark $m_s$ and the QCD coupling constant $\alpha_s$. When $m_s = 0$, we have $k = 1/3$, while for $m_s = 250$ MeV/$c^2$, $k \approx 0.28$. The bag constant $B$ typically lies in the range $0.982B_0 < B < 1.525B_0$ for $m_s = 0$, with $B_0 = 60$ MeV fm$^{-3} $ \cite{N.Stergioulas}. In this work, we choose $B = 60$ MeV fm$^{-3} $, corresponding to $B^{1/4} \approx 147$ MeV and $\alpha_s = 0$, following Figure 1 of Ref. \cite{MIT_jeff_B_value}. Additionally, we work in the ultra-relativistic limit by setting $m_s = 0$, which yields $k = 1/3$.

These equations of state are illustrated in Fig. \ref{fig.eos}, with the transition between the MIT bag model and the unified hadronic equations of state highlighted in the inset.

\section{Initial Conditions and Numerical Integration}\label{ch3_sec_ic}

The modified field equations (\ref{ch3_Ricci_DE}), (\ref{ch3_lambda_DE}), and (\ref{ch3_nu_DE}), together with the Tolman-Oppenheimer-Volkoff (TOV) equation (\ref{ch3_TOV_dP}), form a system of coupled differential equations. These equations are solved numerically, using initial conditions that are consistent with the required asymptotic behavior.

To facilitate numerical integration, we first recast the differential equations (\ref{ch3_Ricci_DE}), (\ref{ch3_lambda_DE}), (\ref{ch3_nu_DE}), and (\ref{ch3_TOV_dP}) into dimensionless form. This is achieved by introducing the following dimensionless variables:
\begin{equation}
\eta = \frac{r}{r_g}, \quad \tilde{P} = \frac{P}{P_0}, \quad \tilde{\rho} = \frac{\rho}{\rho_0}, \quad \chi = R r_g^2, \quad \tilde{T} = \frac{T}{P_0}, \quad \Omega = \frac{d\chi}{d\eta},
\end{equation}
where $P_0 = 4B$, $\rho_0 = P_0 / c^2$, and $r_g = GM_\odot / c^2 = 1.4766 \times 10^5$ cm, which corresponds to half the Schwarzschild radius of the Sun.

To perform the numerical integration, appropriate initial conditions must be specified. At the center of the star ($r = 0$), the cumulative mass satisfies $m(r) = 0$. From the definition of cumulative mass in equation (\ref{ch3_lambdamass}), this implies $\lambda(0) = 0$. Furthermore, since the field equations depend only on the derivative $\nu'(r)$ and not on $\nu(r)$ itself, we can set $\nu(0) = \nu_c$, where $\nu_c$ is an arbitrary constant chosen such that the metric becomes asymptotically flat, i.e., $\nu(\infty) = 0$.

To determine the initial value of the Ricci scalar $R$, we begin with a trial central value $R_c$. This value is varied within the range $0 \leq R_c \leq (3P_c - \rho_c c^2)$, where $P_c$ and $\rho_c$ denote the central pressure and density, respectively. The trial value is adjusted based on whether the resulting numerical solution for $R$ diverges or stabilizes. Through this iterative approach, we refine the initial guess until a stable numerical evolution of $R$ is achieved.

In general relativity, the Ricci scalar $R$ is related to the trace of the energy-momentum tensor by $R = -\kappa T$. Since the energy-momentum tensor vanishes outside the stellar surface ($T = 0$), the Ricci scalar is expected to vanish in the vacuum exterior. However, as shown in equation (\ref{ch3_Ricci_asym1}), the Ricci scalar in this modified theory does not vanish outside the star; instead, it decays exponentially as $r \to \infty$.

As a result, the total mass of the star as measured by a distant observer is given by
\begin{equation}\label{ch3_adm_mass}
M = \frac{c^2}{2G} r_{\infty} \left(1 - e^{-\lambda(r_{\infty})} \right),
\end{equation}
which is greater than the stellar mass evaluated at the surface,
\begin{equation}\label{ch3_stel_mass}
M_s = m(r_s) = \frac{c^2}{2G} r_s \left(1 - e^{-\lambda(r_s)} \right),
\end{equation}
where $r_s$ denotes the stellar radius defined by the condition $P = 0$, and $r_\infty$ is the radial coordinate at which the Ricci scalar effectively vanishes (i.e., $R \approx 0$).

This difference arises from the contribution of the Ricci scalar in the exterior region, which carries gravitational energy. This additional energy contributes to the cumulative mass function $m(r)$ through equations (\ref{ch3_lambdamass}) and (\ref{ch3_lambda_DE}).


\begin{table}[t!]
\centering
\caption{\footnotesize The maximum stable mass $M_{\rm max}$ for different values of the parameter $\alpha$. The corresponding values surface mass $m(r_s)$, stellar radius $r_s$, central value of the curvature scalar $R_c$, and central density $\rho_c$ are also displayed. $M_\odot$ denotes one solar mass.}

\vspace{3mm}

\begin{tabular}{ccccccc}
\hline
Equation of & $\alpha $ & $\rho_c\times10^{15}$ & $R_c$   & $m(r_s)$ & $r_s$ & $M_{\infty}$\\
state       & $r_g^2$   & $g.cm^{-3}$           & $r_g^2$ & $M_\odot$& km    &  $M_\odot$ \\
\hline
            & GR        & 2.39                  & -       & 1.85     & 10.75 & 1.85 \\
            & 1.0       & 2.45                  & 0.016   & 1.78     & 10.81 & 1.88 \\
MIT-SLy     & 3.0       & 2.50                  & 0.010   & 1.73     & 10.89 & 1.91 \\
            & 5.0       & 2.55                  & 0.008   & 1.71     & 10.92 & 1.92 \\
            & 10.0      & 2.55                  & 0.005   & 1.69     & 11.01 & 1.95 \\
\hline
            & GR        & 2.28                  & -       & 1.89     & 10.93 & 1.89 \\
            & 1.0       & 2.35                  & 0.015   & 1.81     & 10.98 & 1.92 \\
MIT-BSk20   & 3.0       & 2.40                  & 0.010   & 1.76     & 11.07 & 1.94 \\
            & 5.0       & 2.40                  & 0.008   & 1.74     & 11.13 & 1.96 \\
            & 10.0      & 2.55                  & 0.005   & 1.72     & 11.14 & 1.98 \\
\hline
            & GR        & 2.03                  & -       & 1.98     & 11.60 & 1.98 \\
            & 1.0       & 2.10                  & 0.015   & 1.90     & 11.65 & 2.00 \\
MIT-BSk21   & 3.0       & 2.15                  & 0.010   & 1.84     & 11.72 & 2.03 \\
            & 5.0       & 2.20                  & 0.008   & 1.82     & 11.75 & 2.04 \\
            & 10.0      & 2.20                  & 0.005   & 1.79     & 11.84 & 2.07 \\
\hline
\end{tabular}\label{tab.ch3_fr_com_EOS}
\end{table}



\begin{figure*}[t]
 \includegraphics[scale=0.3]{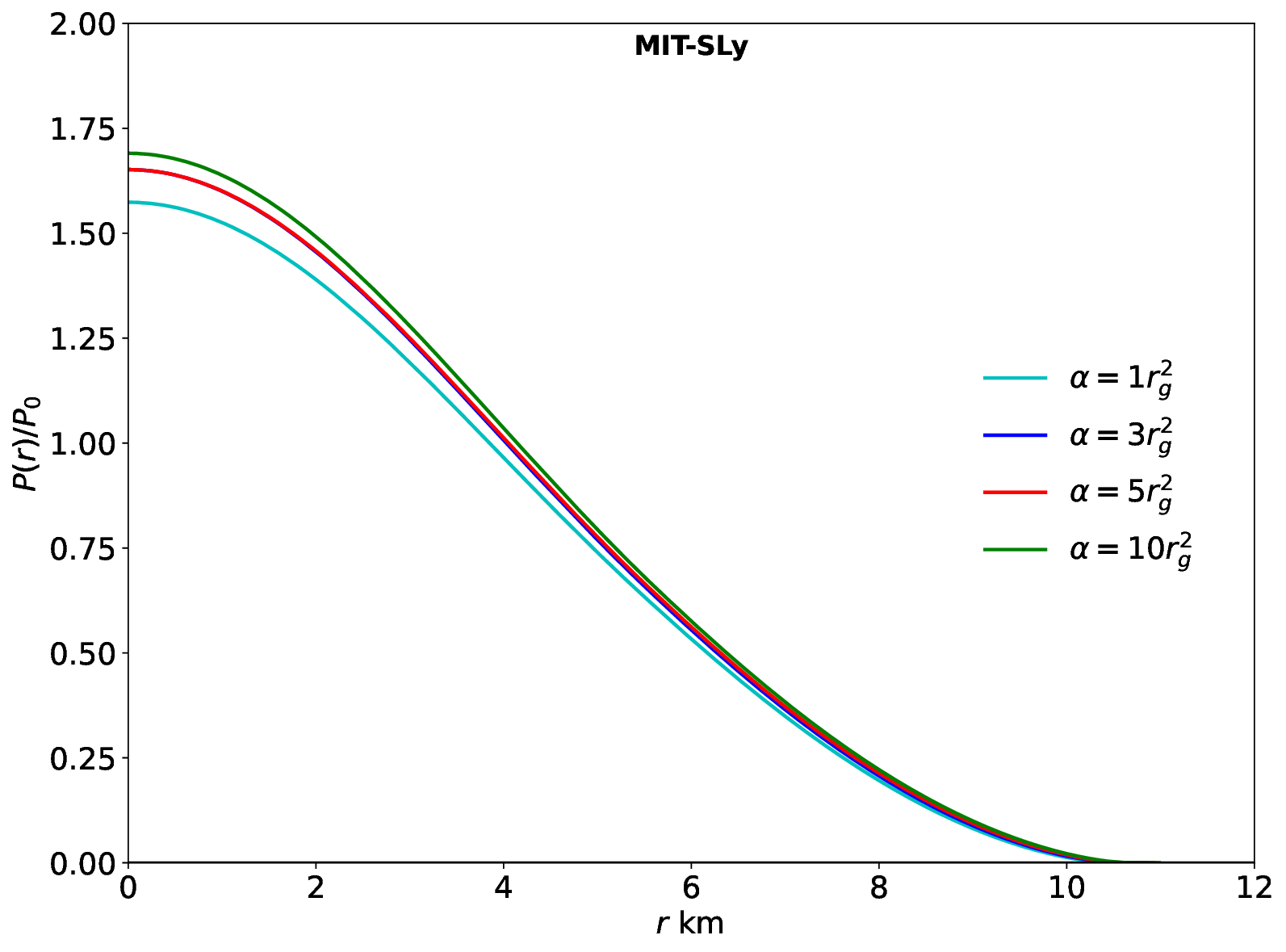}
 \includegraphics[scale=0.3]{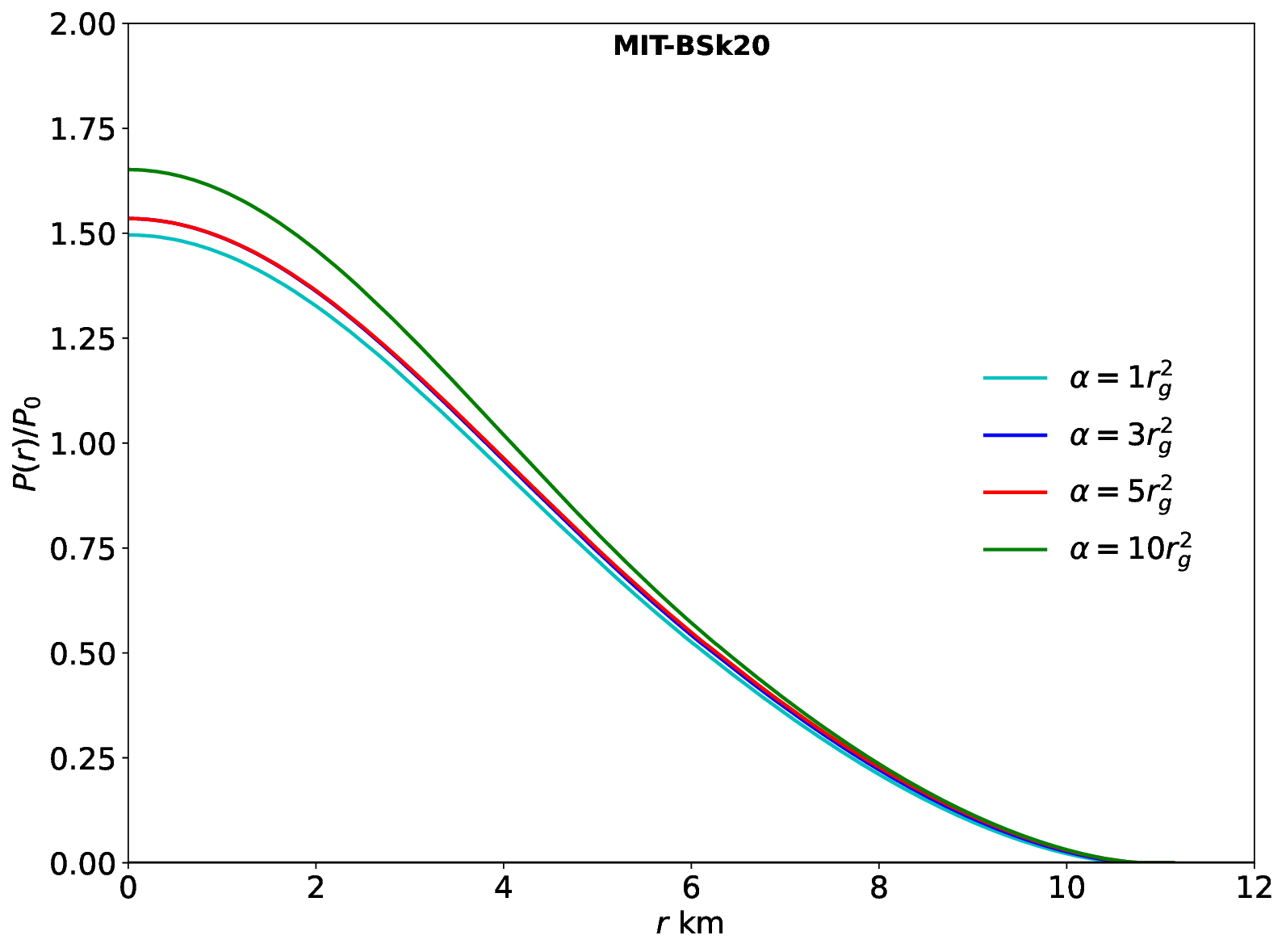}
 \includegraphics[scale=0.3]{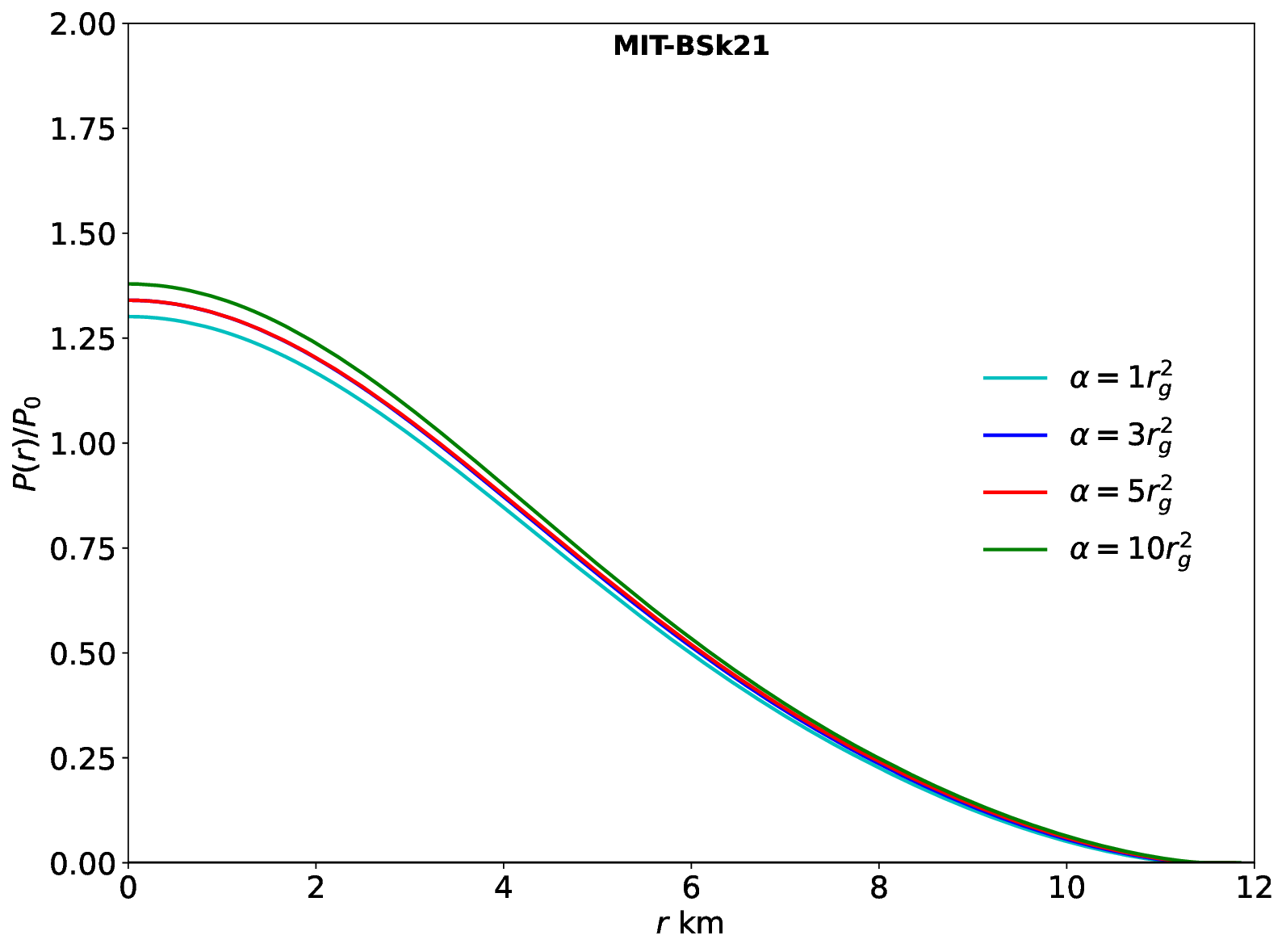}
 \includegraphics[scale=0.3]{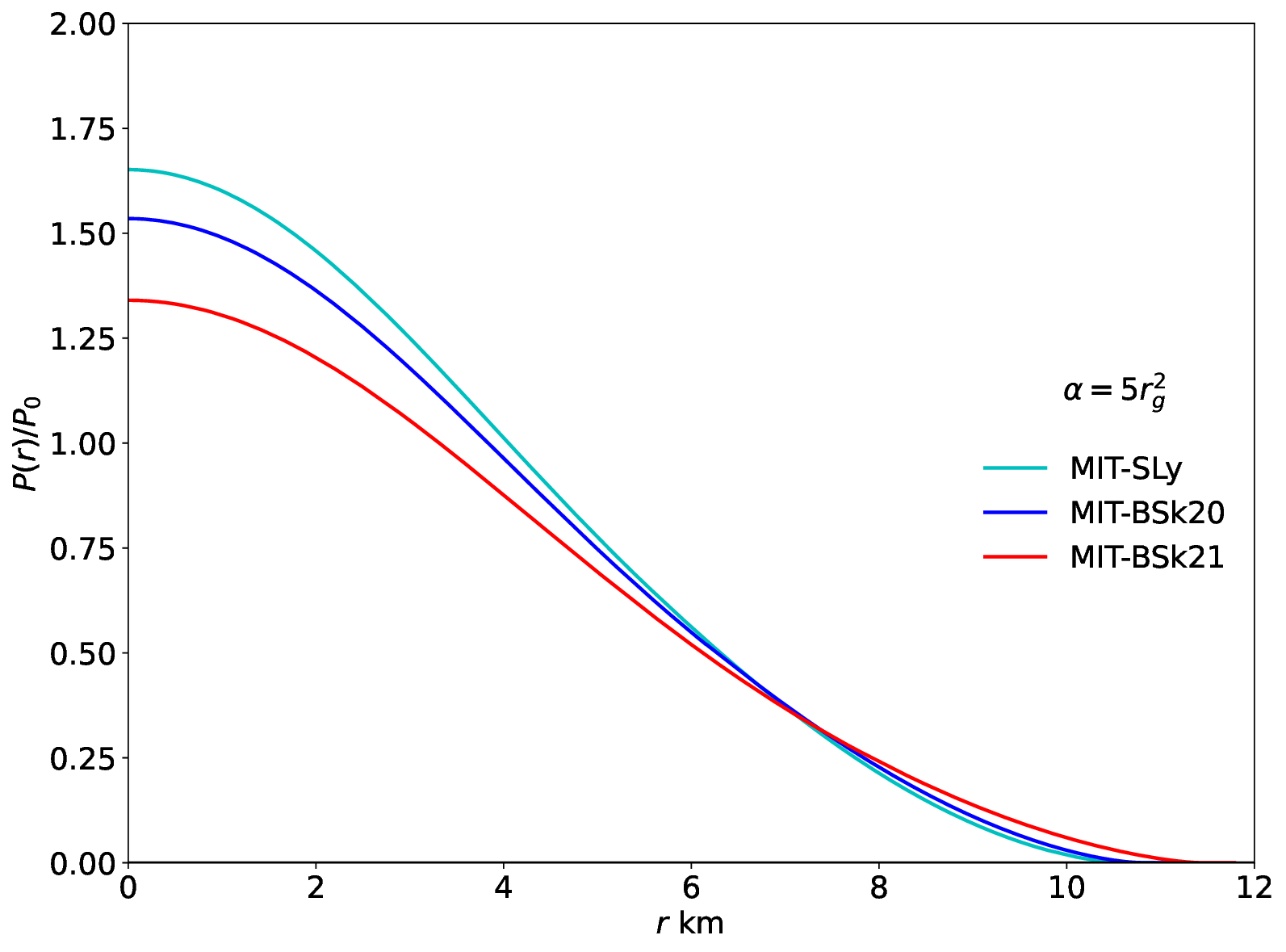}
 \caption{\small The radial pressure profiles of the neutron star with maximal mass (as shown in Table \ref{tab.ch3_fr_com_EOS}) for different values of the Starobinky parameter $\alpha$ with the combined equations of state: MIT-SLy (Top left), MIT-BSk20 (top right), and MIT-BSk21 (bottom left). The bottom-right pannel compares the pressure profiles between the three combined equations of state for $\alpha=5r_g^2$, showing that MIT-SLy can support the highest pressure at the centre.  In all cases, the stellar radius is seen to lie between $10$ km and $12$ km, identified with the pressure dropping to zero value.}\label{ch3_fig_pressure_prof}
\end{figure*}



\begin{figure*}[t]
 \includegraphics[scale=0.3]{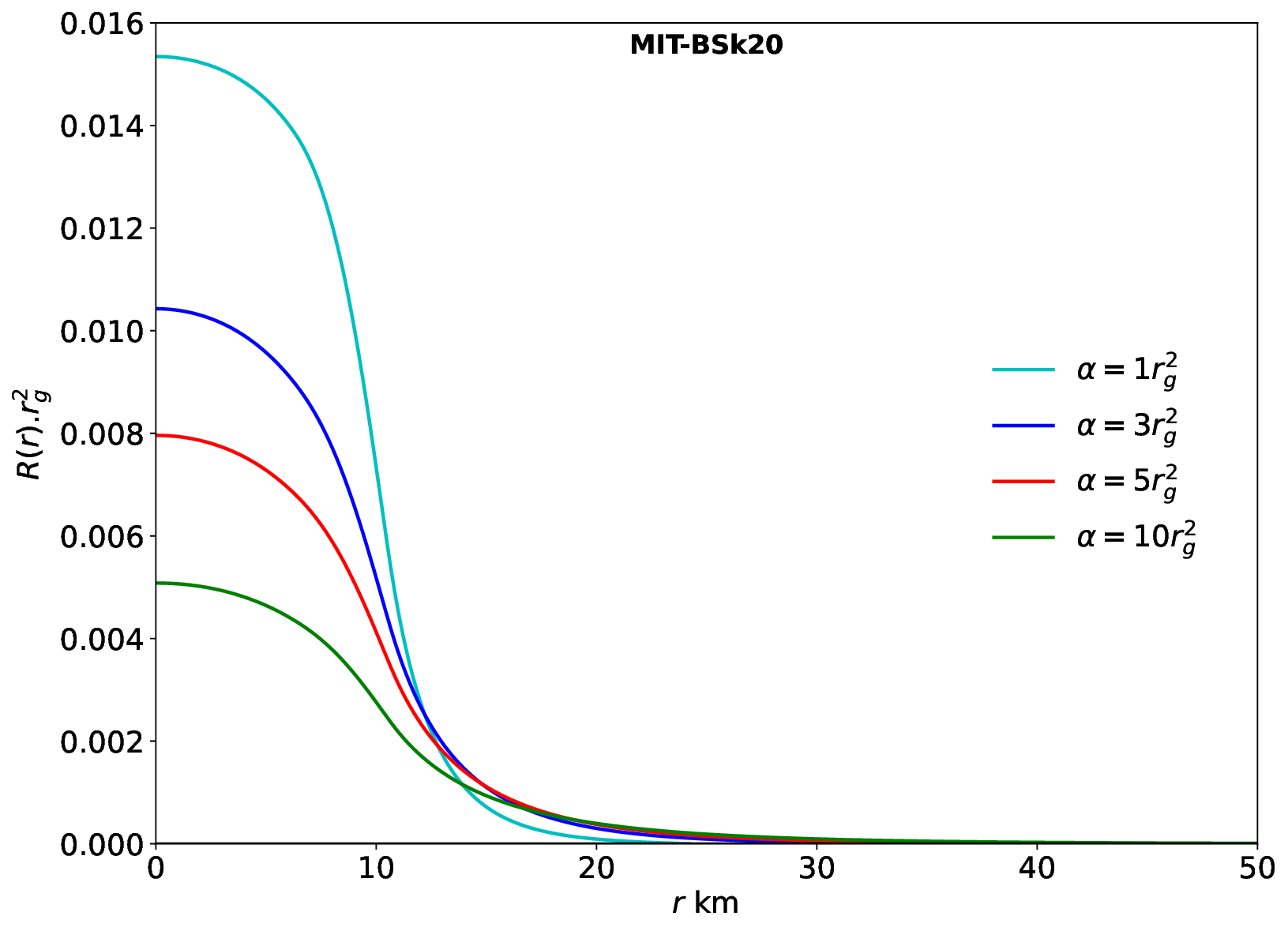}
 \includegraphics[scale=0.3]{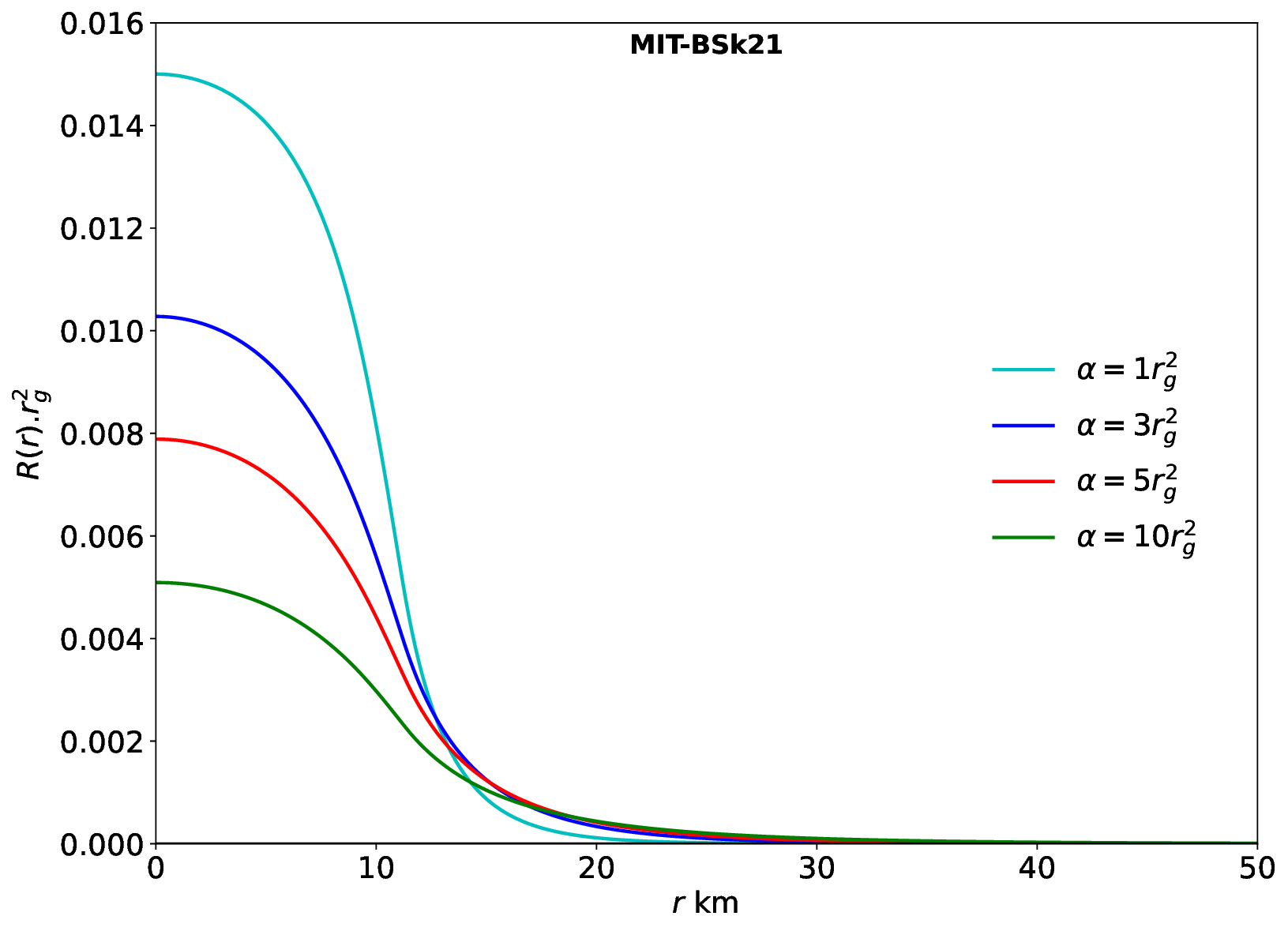}
 \includegraphics[scale=0.3]{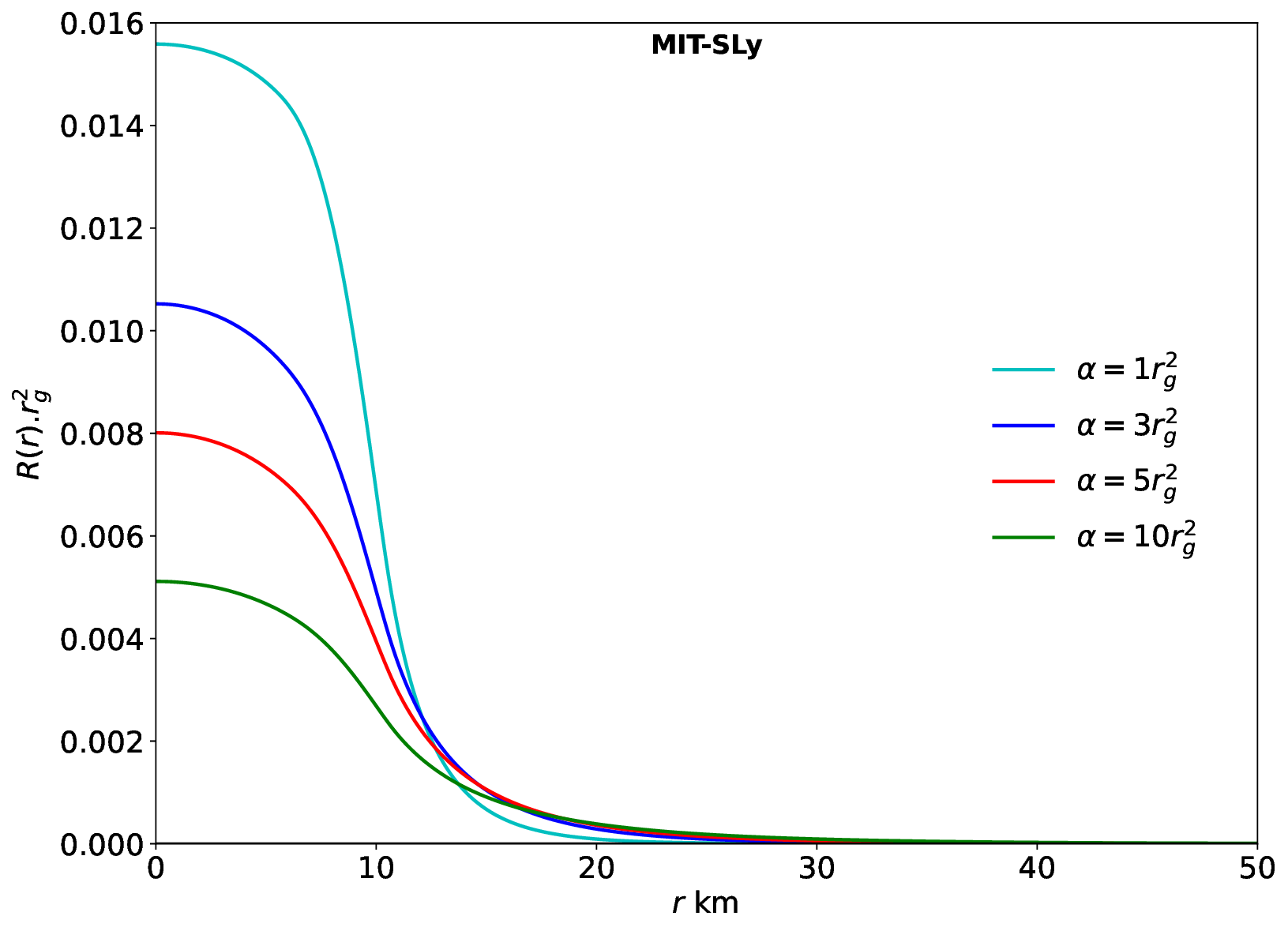}
 \includegraphics[scale=0.3]{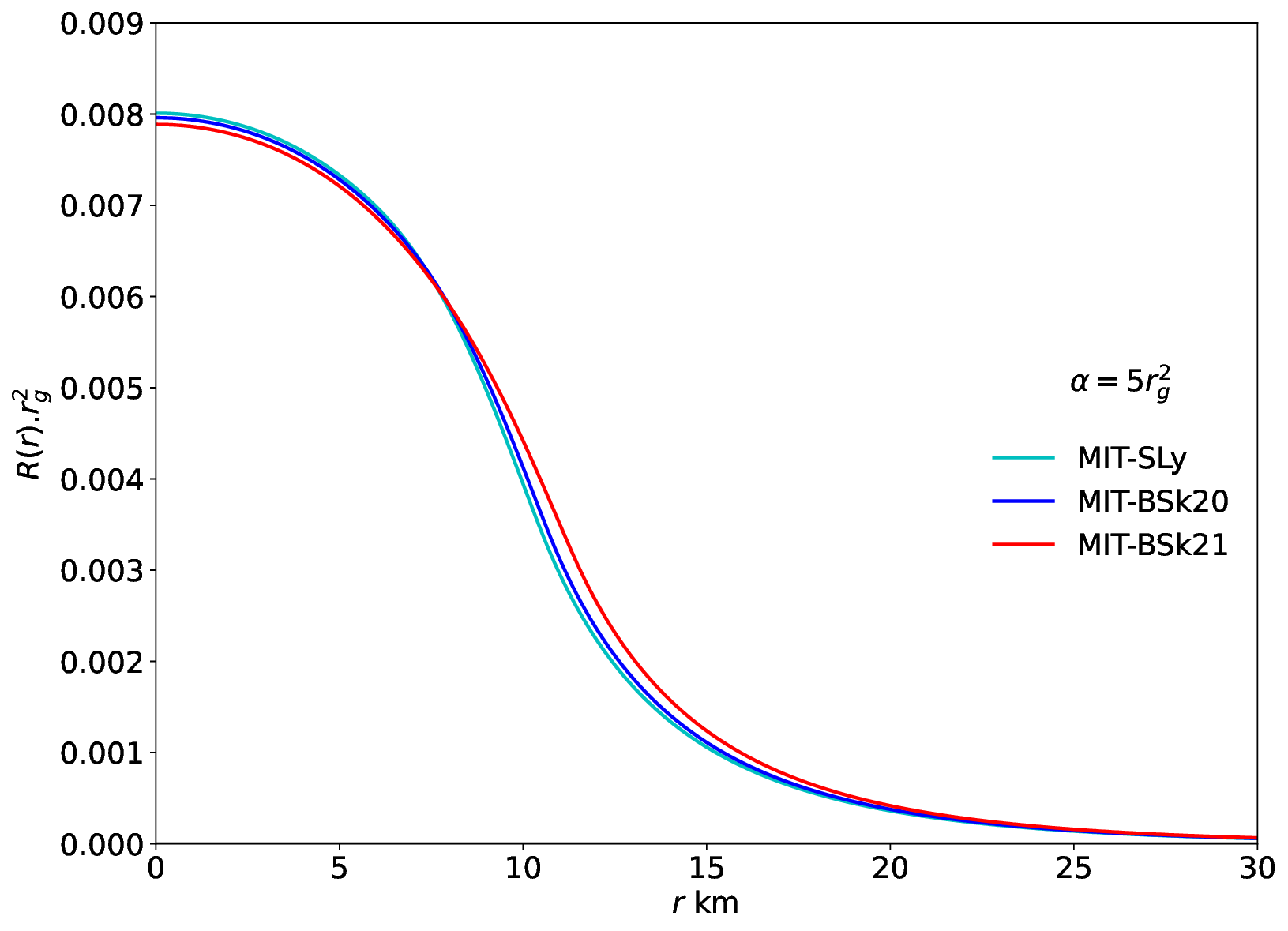}
 \caption{\small The radial profiles of the Ricci scalar $R(r)$ for the neutron star with maximal mass (as shown in Table \ref{tab.ch3_fr_com_EOS}) for different values of the Starobinky parameter $\alpha$ with the combined equations of state: MIT-SLy (Top left), MIT-BSk20 (top right), and MIT-BSk21 (bottom left). The bottom-right pannel compares the profiles between the three combined equations of state for $\alpha=5r_g^2$, showing no remarkable difference in the value of the Ricci scalar at the centre. In all cases, the Ricci scalar does not fall to zero immediately outside the stellar radius; on the contrary, it falls gradually to zero extending beyond $50$ kms.  }\label{ch3_fig_Ricci_prof}
\end{figure*}



\begin{figure*}[t]
 \includegraphics[scale=0.3]{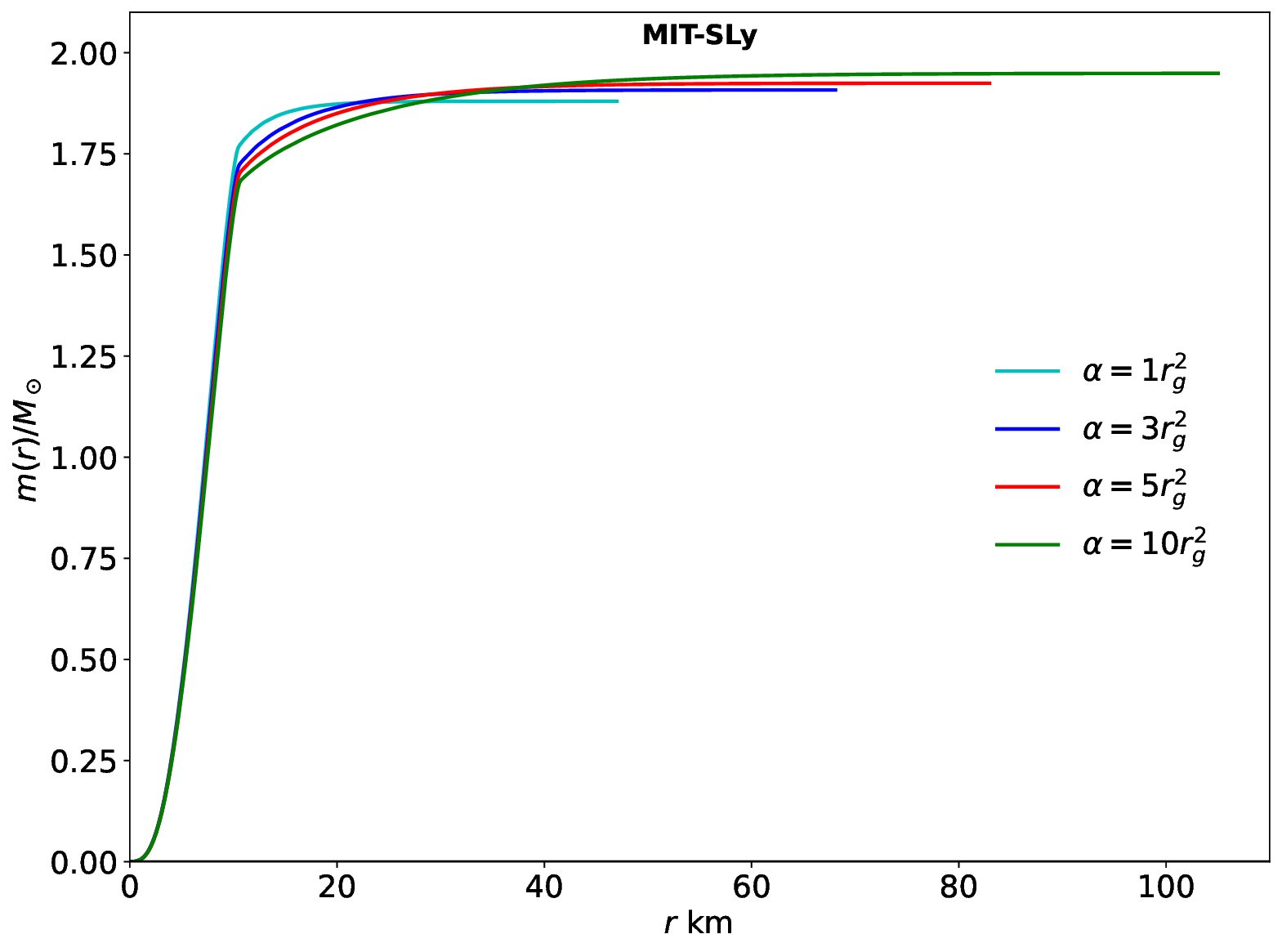}
 \includegraphics[scale=0.3]{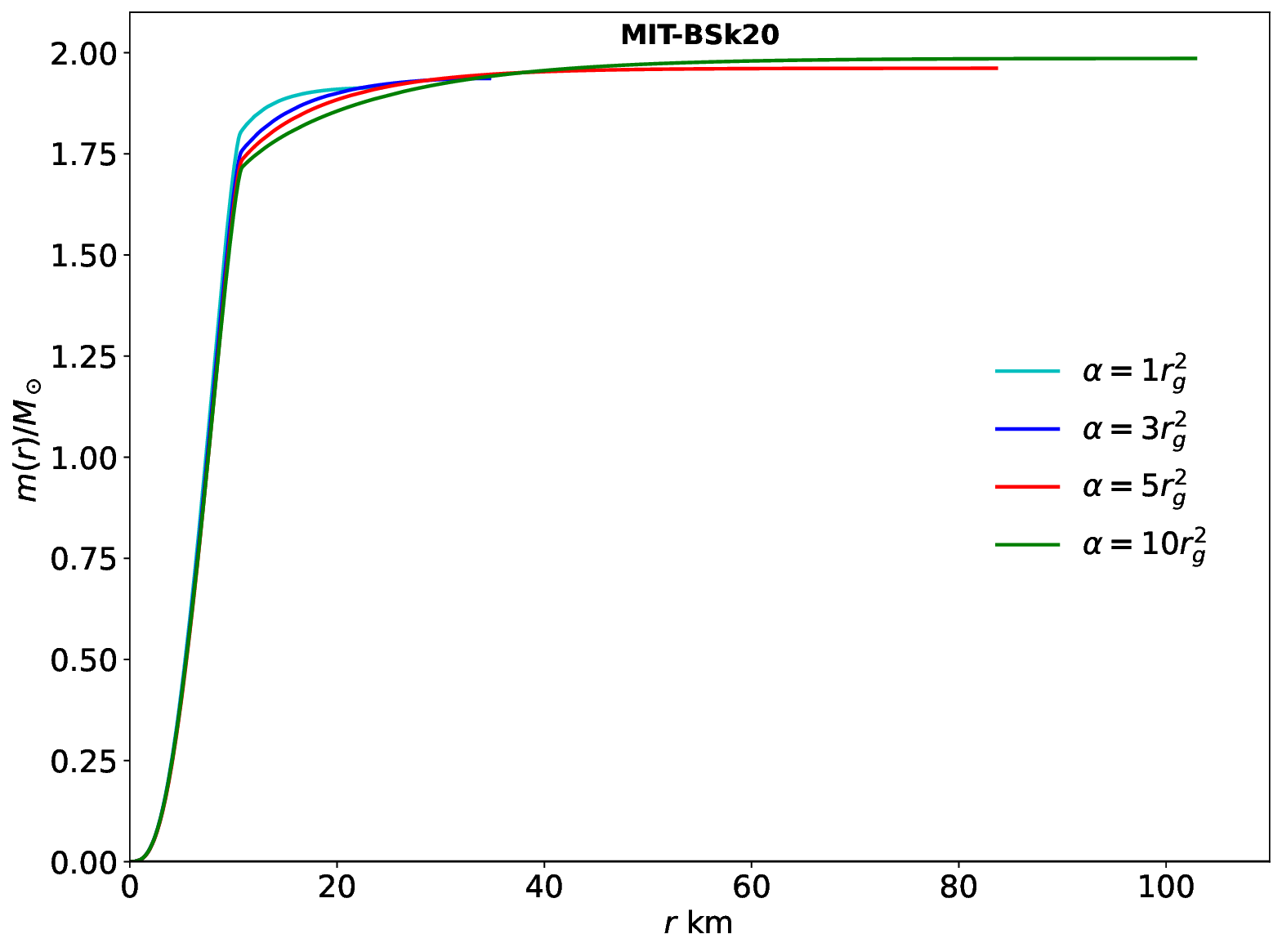}
 \includegraphics[scale=0.3]{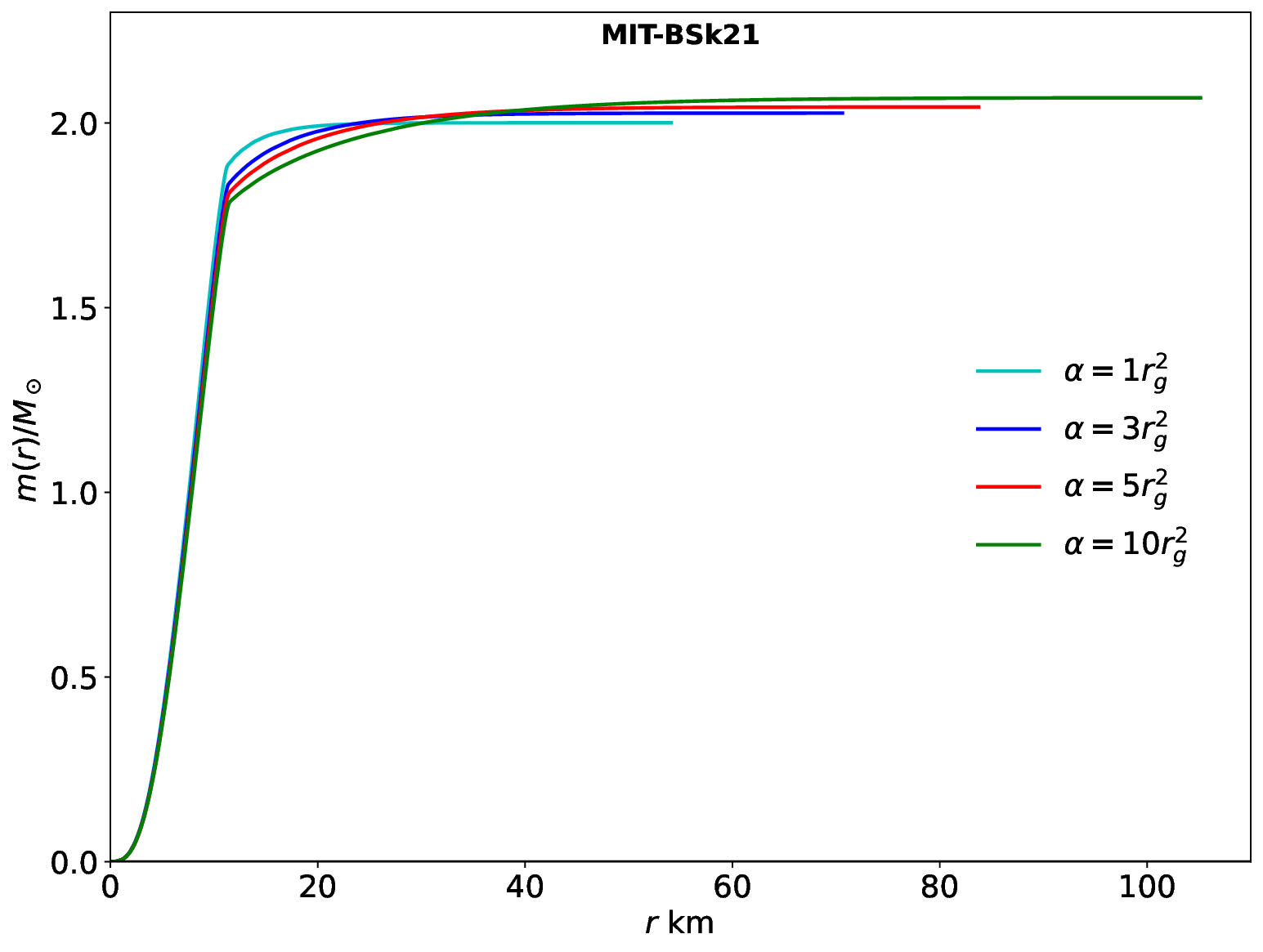}
 \includegraphics[scale=0.3]{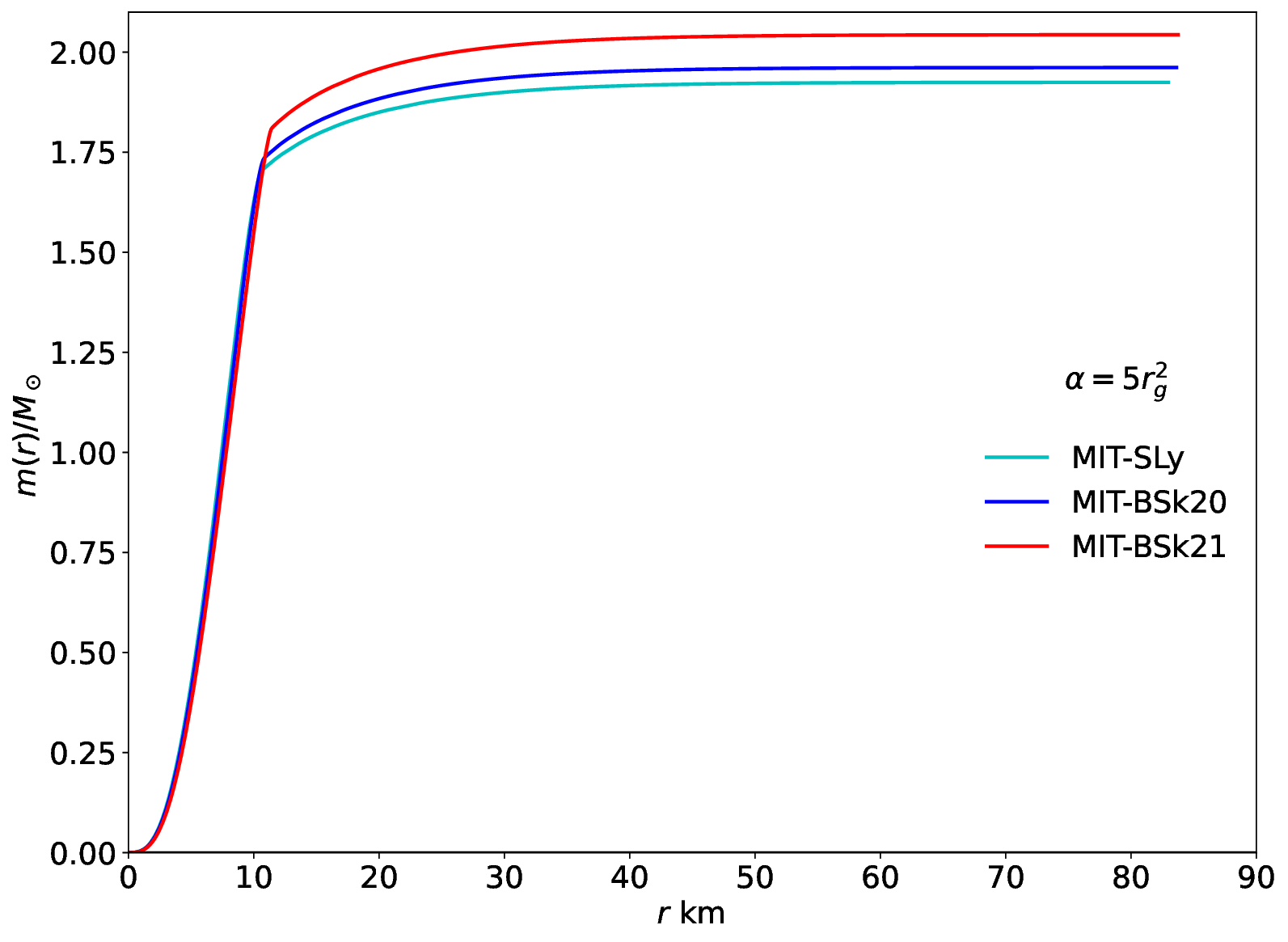}
 \caption{\small The radial profiles of the cumulative mass $m(r)$ for the neutron star with maximal mass (as shown in Table \ref{tab.ch3_fr_com_EOS}) for different values of the Starobinky parameter $\alpha$ with the combined equations of state: MIT-SLy (Top left), MIT-BSk20 (top right), and MIT-BSk21 (bottom left). The bottom-right pannel compares the profiles between the three combined equations of state for $\alpha=5r_g^2$, showing that the ADM mass attains its maximum value for MIT-BSk21 equation of state.  In all cases, the mass value gradually increases to the ADM mass beyond $50$ kms.}\label{ch3_fig_mass_prof}
\end{figure*}


\section{Stellar structure}\label{ch3_result}

In this section, we present the numerical solutions obtained from integrating the modified field equations under various choices of the Starobinsky parameter $\alpha$ and combined equations of state. Our focus is on understanding how these parameters influence the mass-radius relation, pressure distribution, and other internal properties of neutron stars. The results are summarized in tabular and graphical form, highlighting the key physical trends.

Table \ref{tab.ch3_fr_com_EOS} lists the maximum stable mass $M_{\rm max}$ of neutron stars for different values of the Starobinsky parameter $\alpha$. Beyond this threshold, the star becomes unstable and undergoes gravitational collapse. Also shown are the corresponding values of the surface mass $m(r_s)$, stellar radius $r_s$, central curvature scalar $R_c$, and central density $\rho_c$.

From the table, it is evident that the combined equation of state MIT-BSk21 supports a maximum mass of up to $2.07 \, M_\odot$ when $\alpha = 10\,r_g^2$. Furthermore, increasing the value of $\alpha$, the maximum mass can be made consistent with recent observational constraints, such as reported by Cromartie et al. (2019) for the millisecond pulsar MSP J0740+6620, which has a measured mass of $2.14^{+0.10}_{-0.09} \, M_\odot$ \cite{Cromartie2020}.

Figure \ref{ch3_fig_pressure_prof} presents the radial pressure profiles $P(r)$ corresponding to the maximal mass configurations listed in Table \ref{tab.ch3_fr_com_EOS}. Each panel shows the effect of varying $\alpha$ for a specific combination of equations of state: MIT-SLy (top left), MIT-BSk20 (top right), and MIT-BSk21 (bottom left). The bottom-right panel compares the pressure profiles of all three combinations for $\alpha = 5\,r_g^2$, revealing that MIT-SLy produces the highest central pressure. Across all models, the stellar radius is consistently found between 10 km and 12 km, marked by the radius at which the pressure drops to zero.


\begin{figure*}[t]
 \includegraphics[scale=0.3]{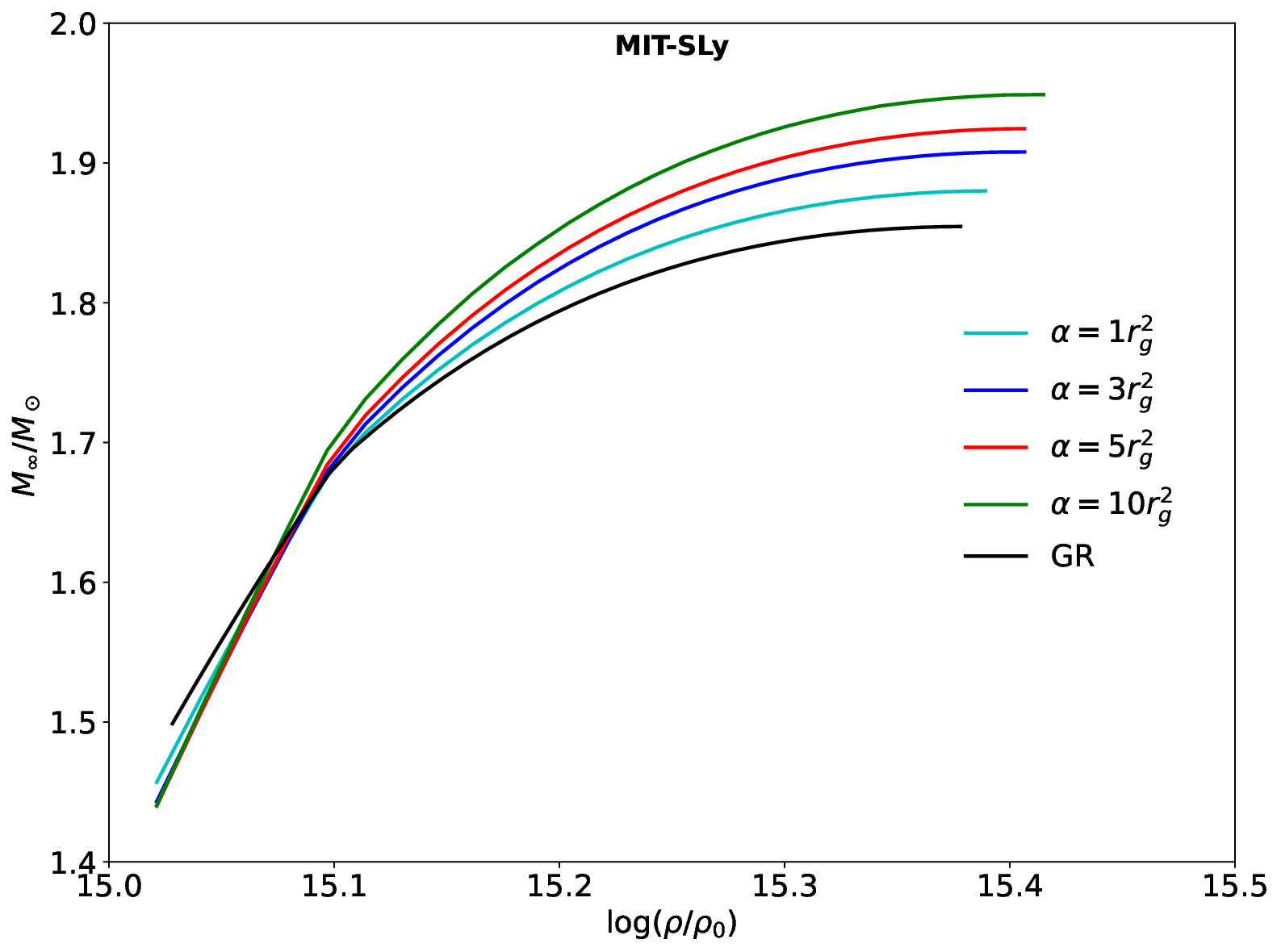}
 \includegraphics[scale=0.3]{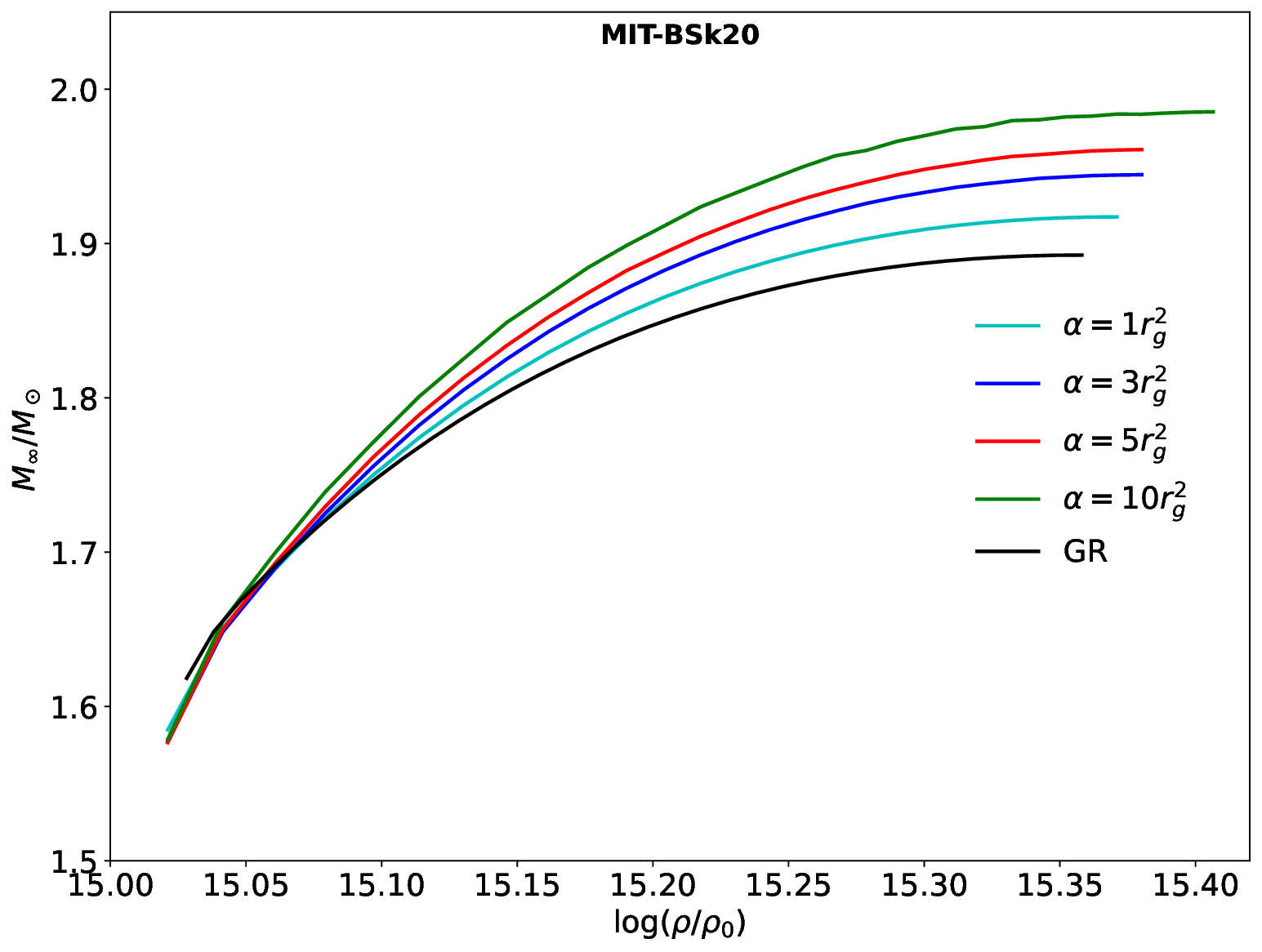}
 \includegraphics[scale=0.3]{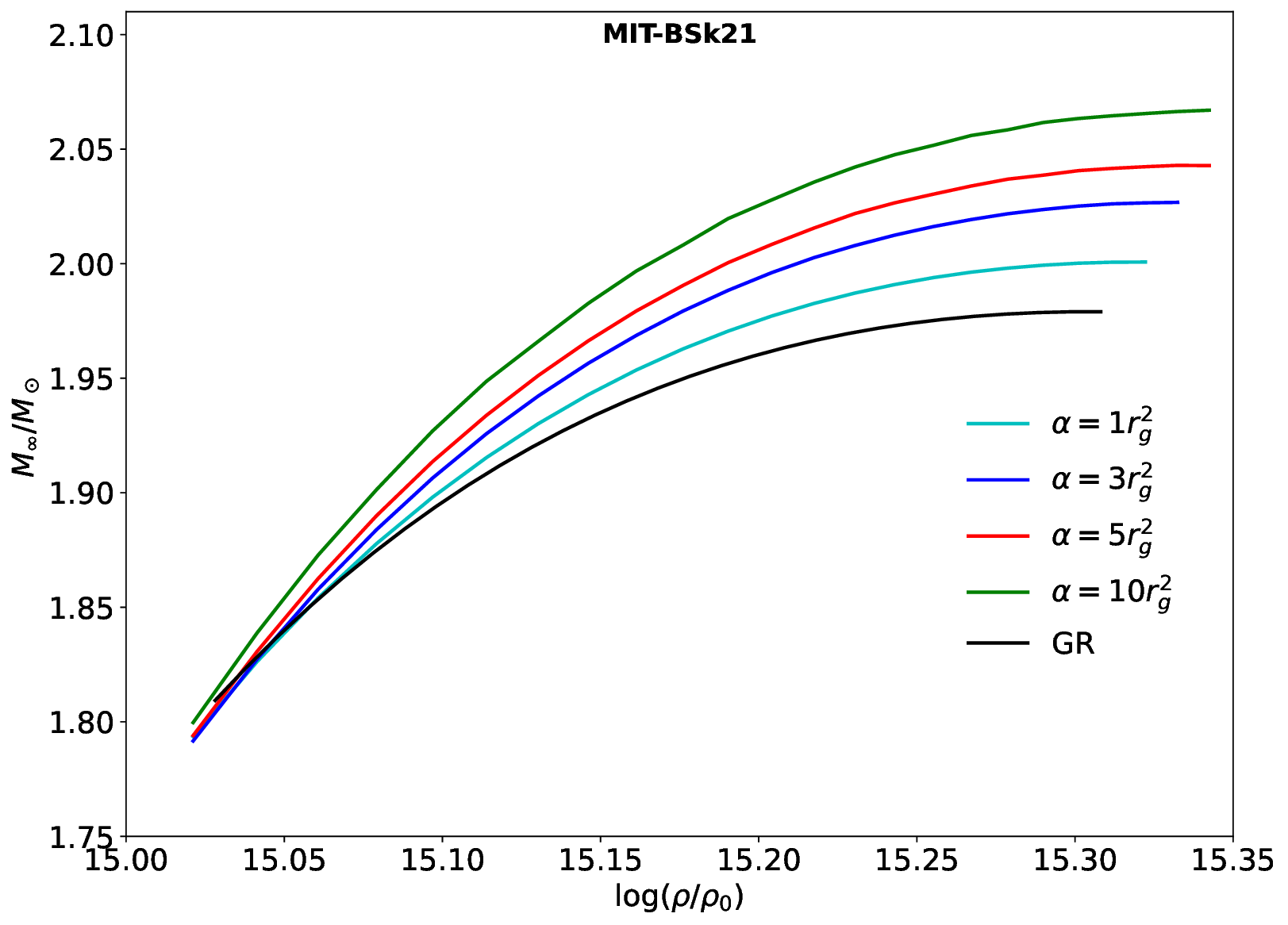}
 \includegraphics[scale=0.3]{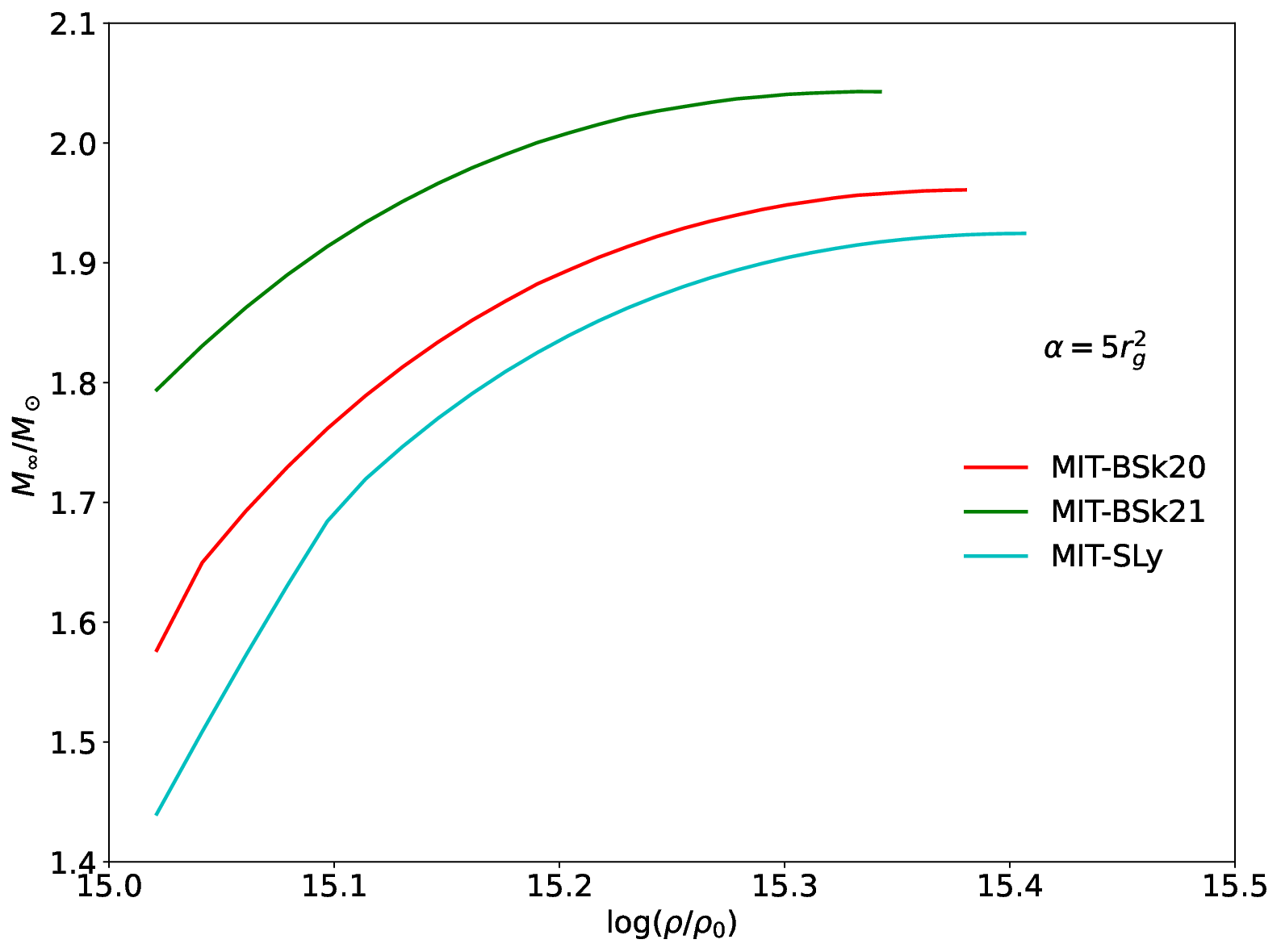}
 \caption{\small ADM mass $M_\infty$ versus central density $\rho_c$ for the neutron star for different values of the Starobinky parameter $\alpha$ with the combined equations of state: MIT-SLy (Top left), MIT-BSk20 (top right), and MIT-BSk21 (bottom left). For comparison, the corresponding graphs with general relativity (GR) are also shown. The bottom-right pannel compares the curves between the three combined equations of state for $\alpha=5r_g^2$.}\label{ch3_fig_mass_dens}
\end{figure*}



\begin{figure*}[t]
 \includegraphics[scale=0.3]{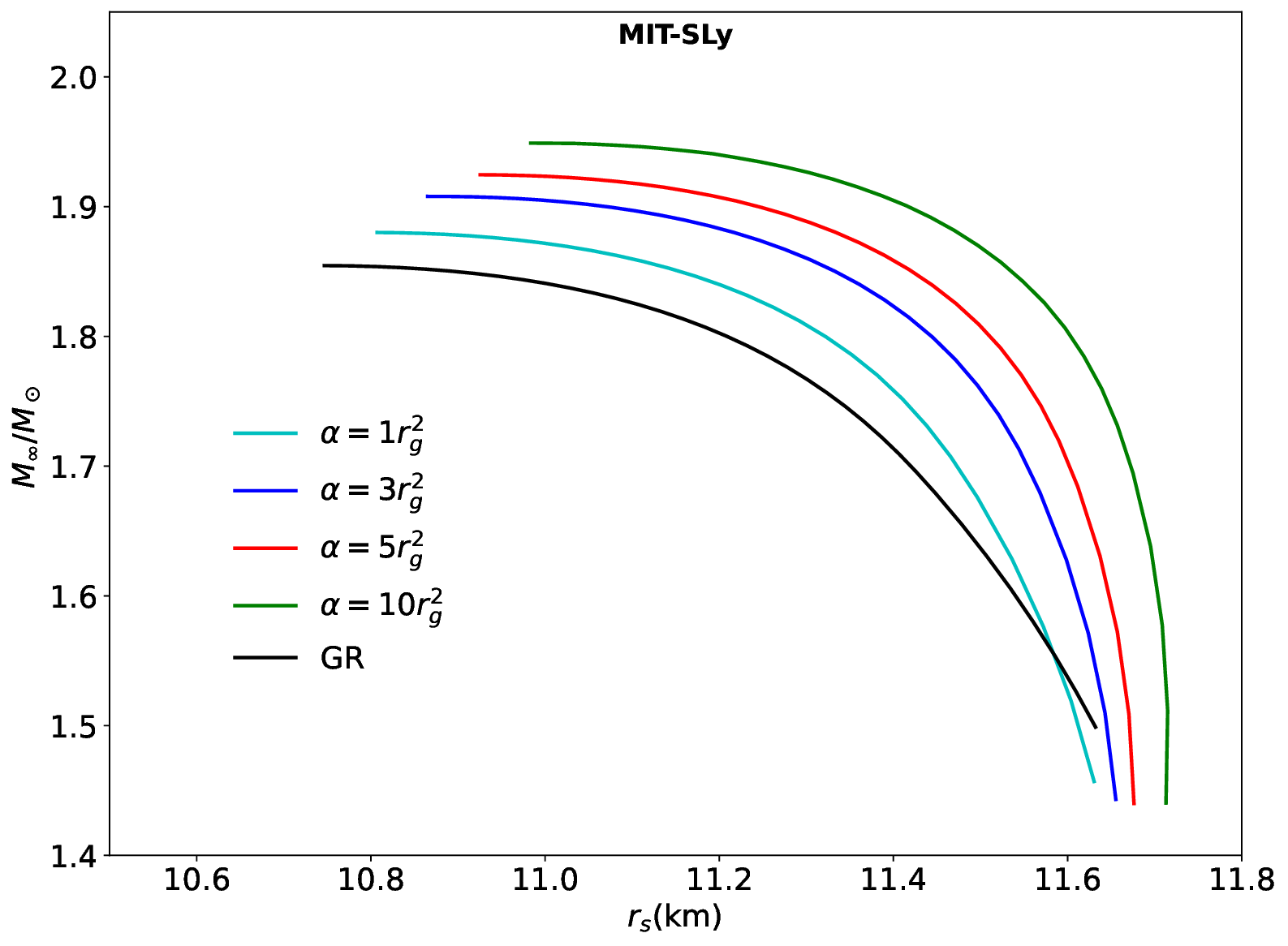}
 \includegraphics[scale=0.3]{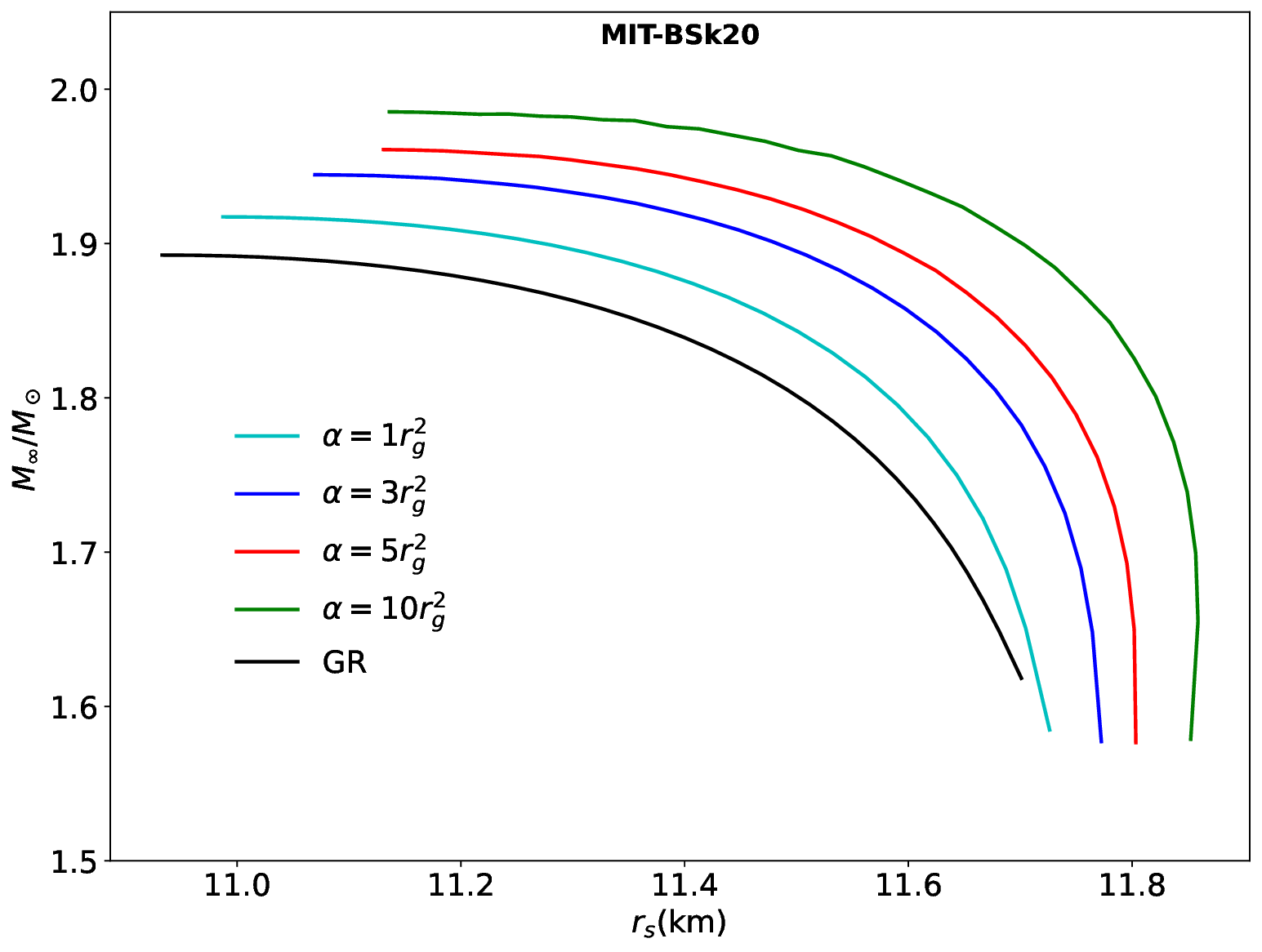}
 \includegraphics[scale=0.3]{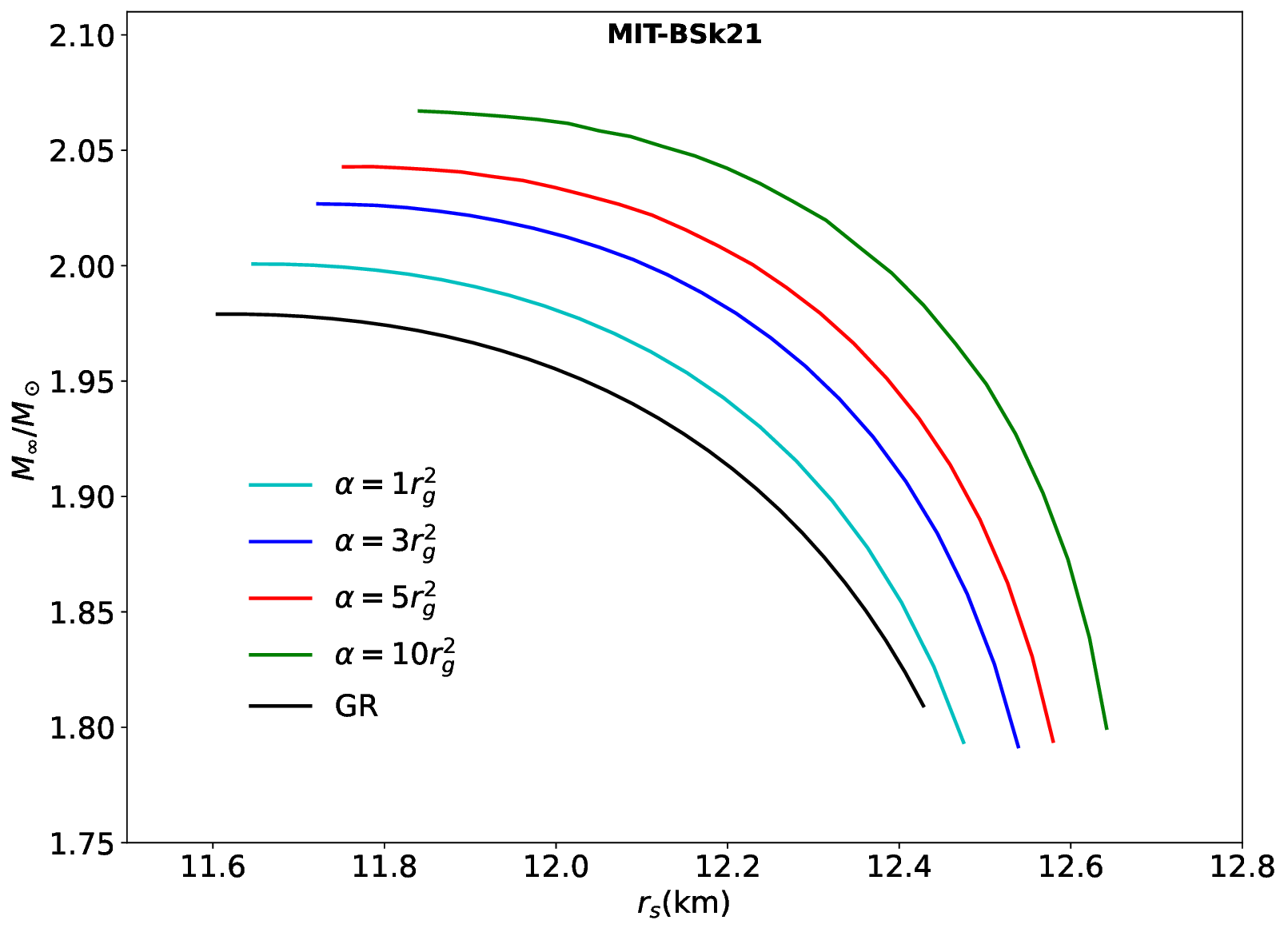}
 \includegraphics[scale=0.3]{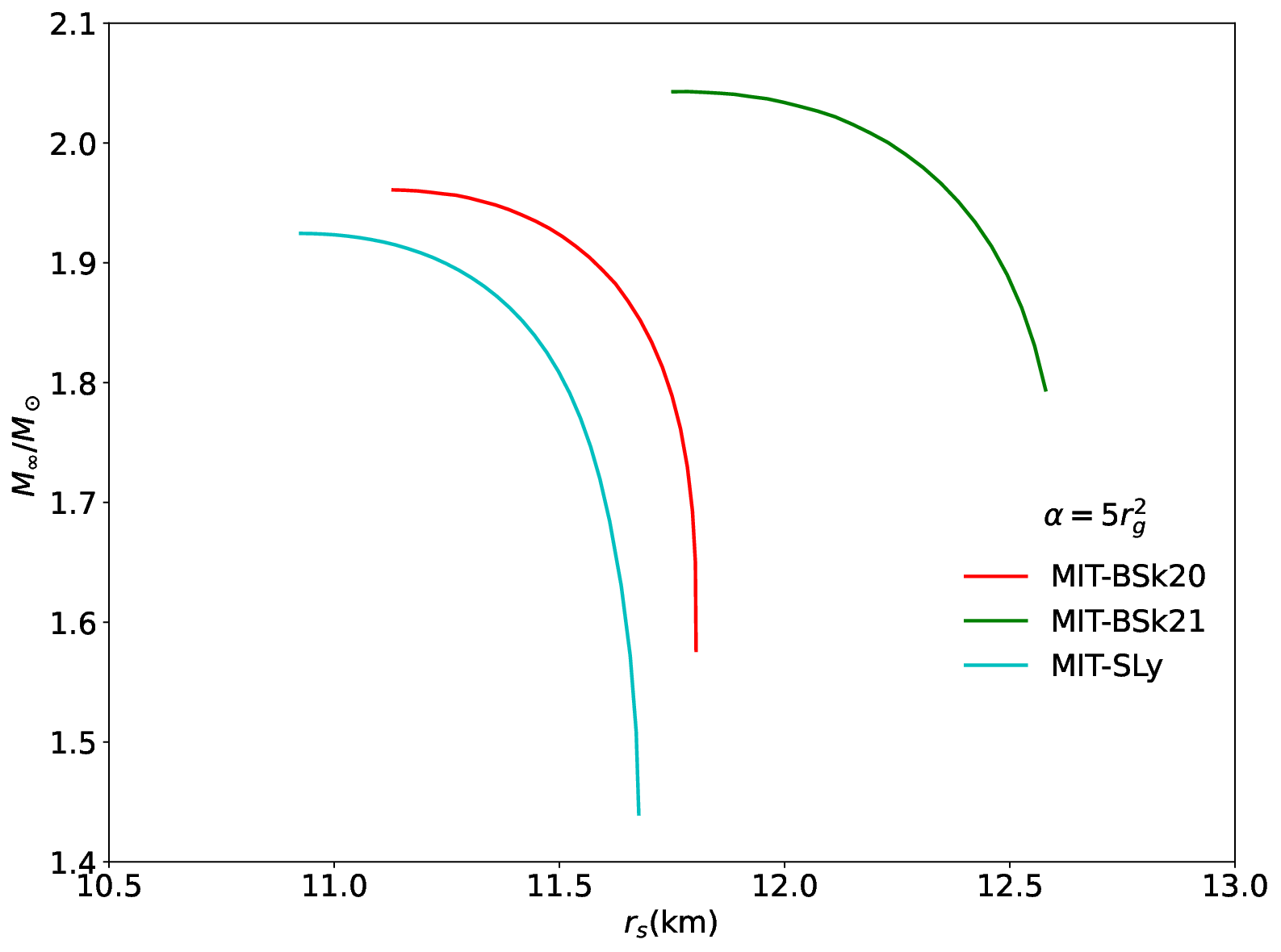}
 \caption{\small ADM mass $M_\infty$ versus stellar radius $r_s$ for the neutron star for different values of the Starobinky parameter $\alpha$ with the combined equations of state: MIT-SLy (Top left), MIT-BSk20 (top right), and MIT-BSk21 (bottom left). For comparison, the corresponding graphs with general relativity (GR) are also shown. The bottom-right pannel compares the curves between the three combined equations of state for $\alpha=5r_g^2$.}\label{ch3_fig_mass_rad}
\end{figure*}



Figure \ref{ch3_fig_Ricci_prof} illustrates the radial profiles of the Ricci scalar $R(r)$ for neutron stars at their respective maximal masses, as listed in Table \ref{tab.ch3_fr_com_EOS}, for various values of the Starobinsky parameter $\alpha$. The panels correspond to different combined equations of state: MIT-SLy (top left), MIT-BSk20 (top right), and MIT-BSk21 (bottom left). The bottom-right panel provides a direct comparison of the Ricci scalar profiles for all three equations of state at $\alpha = 5\,r_g^2$, showing no significant difference in the central value of $R$. In all cases, the Ricci scalar does not drop to zero immediately outside the stellar surface. Instead, it decreases gradually and extends well beyond 50 km from the center. This behavior contrasts with general relativity, where the Ricci scalar vanishes abruptly at the stellar boundary. The non-zero values of the Ricci scalar in the exterior region reflect the presence of an additional scalar degree of freedom---commonly referred to as the scalaron---arising in Starobinsky gravity.

Figure \ref{ch3_fig_mass_prof} shows the radial profiles of the cumulative mass function $m(r)$ for the same set of maximal mass neutron stars. The layout is the same as that of the previous figure, with different panels representing the MIT-SLy (top left), MIT-BSk20 (top right), and MIT-BSk21 (bottom left) combinations. The bottom-right panel compares the mass profiles for $\alpha = 5\,r_g^2$, revealing that the ADM mass reaches its highest value for the MIT-BSk21 equation of state. In all cases, the mass continues to increase gradually beyond the stellar surface, asymptotically approaching the ADM mass at distances beyond 50 km. This extended mass distribution again highlights the influence of the scalaron, which contributes additional gravitational energy in the vacuum region outside the star in $f(R)$ gravity.

Figure \ref{ch3_fig_mass_dens} shows how the ADM mass $M_\infty$ varies with the central density $\rho_c$ for different values of the Starobinsky parameter $\alpha$, using the combined equations of state: MIT-SLy (top left), MIT-BSk20 (top right), and MIT-BSk21 (bottom left). For comparison, the corresponding curves obtained with general relativity (GR) are also included. The bottom-right panel compares the mass-central density relations among the three equations of state for a fixed value $\alpha = 5\,r_g^2$. In all cases, the ADM mass increases monotonically with the central density, indicating that the Starobinsky gravity model exhibits a physically consistent behavior.

Figure \ref{ch3_fig_mass_rad} presents the relationship between the ADM mass $M_\infty$ and the stellar radius $r_s$ for different values of the Starobinsky parameter $\alpha$, again using the three combined equations of state: MIT-SLy (top left), MIT-BSk20 (top right), and MIT-BSk21 (bottom left). The corresponding general relativity results are provided for comparison. The bottom-right panel compares the three equations of state for $\alpha = 5\,r_g^2$. In every case, the stellar radius decreases monotonically as the ADM mass increases---a behavior expected due to the stronger gravitational compression in more massive neutron stars.

The numerical results confirm that the Starobinsky model of $f(R)$ gravity can support stable neutron star configurations with masses exceeding the general relativistic predictions, depending on the choice of the parameter $\alpha$ and the underlying equation of state. The model produces physically reasonable profiles for pressure, curvature, and mass, and can be made consistent with observational constraints such as the mass of MSP J0740+6620. The extended mass and curvature profiles outside the stellar surface highlight the influence of the scalar degree of freedom introduced in this modified theory of gravity.

\section{Conclusions}\label{ch3_conc}

In this study, we explored the stellar structure of neutron stars within the framework of Starobinsky gravity, defined by the extended gravitational action $f(R) = R + \alpha R^2$. By deriving the modified Tolman-Oppenheimer-Volkoff (TOV) equations for this theory and solving them numerically for various values of the Starobinsky parameter $\alpha$, we examined the effect of the quadratic curvature correction on the physical properties of neutron stars.

Our numerical results, presented in Table \ref{tab.ch3_fr_com_EOS}, demonstrate that the maximum stable mass $M_{\rm max}$ of neutron stars increases with higher values of $\alpha$. This enhancement arises from the additional gravitational support against collapse provided by the higher-order curvature term. Specifically, for the combined MIT-BSk21 equation of state, a maximum mass of $2.07 \, M_\odot$ is attained when $\alpha = 10\,r_g^2$. This suggests that the theoretical prediction can be brought into agreement with the observed mass of the heavy millisecond pulsar MSP J0740+6620 ($2.14^{+0.10}_{-0.09}$ M$_\odot$) reported by Cromartie et al.~\cite{Cromartie2020}.

Nevertheless, previous studies \cite{20Percent_2, 20Percent_1} have shown that rapid rotation can increase the maximum mass of a neutron star by approximately 20--25\% compared to its non-rotating counterpart. Based on this, the static maximal mass of $2.07 \, M_\odot$ obtained in our model with $\alpha = 10\,r_g^2$ suggests that a rotating configuration could reach mass thresholds in the range of $2.48$ to $2.59 \, M_\odot$, further aligning with observational constraints for highly massive neutron stars.

In addition, Figure \ref{ch3_fig_pressure_prof} displays the radial pressure profiles $P(r)$ for various combinations of equations of state and values of the Starobinsky parameter $\alpha$. The results indicate that all modeled neutron stars have radii between 10 and 12 km, which aligns well with current observational constraints. Among the considered models, the MIT-SLy equation of state yields the highest central pressure, emphasizing the significant influence of the chosen nuclear matter model on the internal structure and compactness of the star.

Figure \ref{ch3_fig_Ricci_prof} shows the radial profiles of the Ricci scalar $R(r)$ corresponding to the maximum mass configurations. In contrast to general relativity, where the Ricci scalar vanishes sharply outside the stellar surface, Starobinsky gravity produces a gradual decay of $R(r)$, which remains non-zero well beyond the stellar radius, extending over distances exceeding 50 km. This extended curvature profile is a distinctive feature of $f(R)$ theories, reflecting the presence of an additional scalar degree of freedom---the scalaron. Notably, for a fixed value of $\alpha$, the central values of the Ricci scalar are found to be nearly identical across different equations of state, suggesting a degree of universality in the core curvature behavior within the Starobinsky framework.

Figure \ref{ch3_fig_mass_prof} illustrates the cumulative mass profiles $m(r)$, showing how the total gravitational mass builds up as a function of radius for different values of $\alpha$ and equations of state. In all cases, the mass does not reach its final value at the stellar surface but continues to increase gradually beyond it, approaching the ADM mass only at distances exceeding 50 km. This extended mass accumulation reflects the influence of the scalaron field outside the star---a phenomenon absent in general relativity. Among the models considered, the MIT-BSk21 equation of state yields the highest ADM mass for $\alpha = 5\,r_g^2$, as evident from the bottom-right panel of the figure. These findings highlight the significant role of the additional scalar degree of freedom in shaping both the internal and external mass distribution of neutron stars.

Figure \ref{ch3_fig_mass_dens} examines the relationship between the ADM mass $M_\infty$ and central density $\rho_c$ for various equations of state and values of $\alpha$, alongside the corresponding predictions from general relativity. In every case, $M_\infty$ increases monotonically with central density, consistent with the expected behavior of stable compact stars. This result confirms that Starobinsky gravity maintains the core thermodynamic consistency of stellar models while introducing distinct modifications to the exterior mass distribution through the scalaron contribution.

Finally, Figure \ref{ch3_fig_mass_rad} presents the ADM mass-radius relation for different values of $\alpha$ and equations of state, with comparisons to general relativistic predictions. As expected, the stellar radius decreases monotonically with increasing mass, indicating that more massive stars are more compact due to stronger gravitational binding. This behavior demonstrates that the inclusion of the quadratic curvature term in the gravitational action does not violate the physical intuition of the mass-radius relation. Instead, it supports the consistency and robustness of Starobinsky gravity as a viable extension of general relativity in the high-density regime.

In summary, our analysis demonstrates that Starobinsky gravity, through the inclusion of a quadratic curvature correction, offers a consistent and physically viable extension of general relativity for modeling neutron stars. The theory not only supports higher maximal masses---compatible with recent astrophysical observations---but also introduces distinctive features in the curvature and mass profiles outside the stellar surface, attributable to the scalaron field. Importantly, key physical relationships, such as the mass-central density and mass-radius curves, remain preserved. These results highlight the potential of modified gravity theories to address open questions in the physics of compact objects while remaining consistent with observational constraints.

\section*{Acknowledgments}
Muhammed Shafeeque is supported through a Research Fellowship by the Ministry of Education, Government of India. The Authors would like to thank the Indian Institute of Technology Guwahati for providing access to computing and supercomputing facilities.


\end{document}